\documentclass[useAMS,usenatbib]{mn2e}

\usepackage{amssymb}
\usepackage{gensymb}
\usepackage{graphicx}
\usepackage{subfig}

\title[NIR HST polarimetry of FRII NLRG] {Near-infrared {\em Hubble Space Telescope} polarimetry of a complete sample of narrow-line radio galaxies}
\author[Ram\'irez et al.]{E. A. Ram\'irez$^{1}$\thanks{E-mail: e.ramirez@usp.br}, C. N. Tadhunter$^{2}$, D. Axon$^{3,4,}$\thanks{Deceased.}, D. Batcheldor$^{5}$, \newauthor  C. Packham$^{6}$, E. Lopez-Rodriguez$^{6}$, W. Sparks$^{7}$ and S. Young$^{8}$\\
$^{1}$Universidade de S\~ao Paulo, IAG, Rua do Mat\~ao 1226, Cidade Universit\'aria, S\~ao Paulo 05508-900, Brazil.\\
$^{2}$Department of Physics and Astronomy, University of Sheffield, Sheffield S3 7RH, UK.\\
$^{3}$Physics Department, Rochester Institute of Technology, Rochester, NY 14623, USA.\\
$^{4}$School of Mathematical and Physical Sciences, University of Sussex, Brighton BN1 9QH, UK.\\
$^{5}$Department of Physics and Space Sciences, Florida Institute of Technology, FL 32901, USA.\\
$^{6}$Department of Physics \& Astronomy, University of Texas at San Antonio, One UTSA Circle, San Antonio, TX 78249, USA.\\
$^{7}$SpaceTelescope Science Institute, 3700 San Martin Drive, Baltimore, MD21218, USA.\\
$^{8}$Centre for Astrophysics Research, Science and Technology Research Institute, University of Hertfordshire, Hatfield AL10 9AB, UK.
 }

\begin{document}

\date{Accepted 2014 July 7.  Received 2014 June 30; in original form 2014 April 10}

\pagerange{\pageref{firstpage}--\pageref{lastpage}} \pubyear{2014}

\maketitle

\label{firstpage}

\begin{abstract}

We present an analysis of 2.05~$\mu$m {\em Hubble Space Telescope}  ({\em HST}) polarimetric data for a sample of 13 nearby Fanaroff-Riley type II (FRII) 3CR radio sources ($0.03<z<0.11$) that are classified as narrow line radio galaxies (NLRG) at optical wavelengths. We find that  the  compact cores of the NLRG in our sample are intrinsically highly polarised in the near-IR ($6 < P_{2.05\mu m} < 60$ per cent), with the electric-vector (E-vector) perpendicular to the radio axis in 54 per cent of the sources. The levels of extinction required to produce near-infrared polarisation by the dichroic extinction mechanism are consistent with the measured values reported in \citet{Ramirez:2014}, provided that this mechanism has its maximum efficiency. This consistency suggests that the nuclear polarisation could be due to dichroic extinction. In this case, toroidal magnetic fields that are highly coherent would be required in the circumnuclear tori to align the elongated dust grains responsible for the dichroic extinction. However, it is not entirely possible to rule out  other polarisation mechanisms (e.g. scattering, synchrotron emission) with our observations at only one near-IR wavelength. Therefore further polarimetry observations at mid-IR and radio wavelengths will be required to test whether all the near-IR polarisation is due to dichroic extinction.

\end{abstract}


\section{Introduction}

The orientation-based unified schemes for active galactic nuclei \citep[AGN.][]{Antonucci:1993,Urry:1995} propose that the quasar continuum light in type-2 radio galaxies is hidden from our direct view by an obscuring torus at optical wavelengths. 

In the case of powerful radio galaxies (PRG), polarimetry observations provide strong evidence to support the unified schemes. For instance, optical spectropolarimetric observations of narrow-line radio galaxies (NLRG) show broad H$\alpha$ lines detected in a significant number of objects \citep{Antonucci:1984,Tran:1995,Goodrich:1992,Ogle:1997,Cohen:1999}. These detections are attributed to light from the broad-line region (BLR) scattered by material out of the plane of the torus.

Optical polarimetric imaging of Fanaroff and Riley type II \citep[FRII;][]{Fanaroff:1974} PRG also shows a centro-symmetric pattern of polarisation vectors contained within a bi-cone structure in seven out of eight NLRG in the \citet{Cohen:1999} sample, including Cygnus~A \citep[see also][]{Tadhunter:1990, Ogle:1997}. This, together with the detection of broad polarised H$\alpha$ lines, reveals the presence of  broad-cone reflection nebulae illuminated by the AGN.

The direct detection of the unresolved AGN in NLRG is difficult at optical wavelengths, given the high extinction and the evident presence of dust \citep{de_Koff:1996, Chiaberge:1999,Chiaberge:2000,Chiaberge:2002b}. However,  {\em Hubble Space Telescope}  ({\em HST}) observations of Cygnus~A show a clear progression from a double V-shape structure detected at optical and shorter near-IR wavelengths, to  a compact nuclear source at longer near-IR  wavelengths \citep{Djorgovski:1991, Tadhunter:1999}.  More recently, we detected a point-like nucleus at 2.05~$\mu$m in 80 per cent of a nearby sample of NLRG \citep{Ramirez:2014}, suggesting that the central engine is shining directly through a dusty torus in these sources, providing evidence for the unified schemes.

It is not only infrared (IR) imaging and spectroscopic observations that are useful in the study of the central obscuring region in PRG: the analysis of near-IR polarimetric data also has the potential to provide important complementary information on the geometry and intrinsic properties of the central regions. Studies at the $K$-band of the nuclear polarisation of Centaurus~A and Cygnus~A  show high core polarisations of $9\pm1$ per cent and $28\pm1$ per cent, respectively, with the electric-vector (E-vector) position angle ($\textrm{PA}$) aligned perpendicular to the radio jet \citep{Bailey:1986, Tadhunter:2000a}. A similar configuration is found in 3C~223.1 and 3C~433 at 2.2~$\mu$m and 2.05~$\mu$m, respectively: polarisation E-vectors perpendicular to the radio jet axis,  in addition to  dust lanes orientated perpendicular to the radio jets \citep{Antonucci:1990, Ramirez:2009}. This is consistent with a dichroic\footnote[1]{Dichroic extinction is the selective attenuation of light passing through a medium that contains dust grains aligned by a magnetic field.} or scattering origin for the polarisation, since such mechanisms can produce polarisation $\textrm{PA}$s that are perpendicular to the radio jet orientation. In contrast, the polarisation $\textrm{PA}$s for any synchrotron emission from the inner jets tend to be parallel to the radio jet orientation \citep[e.g.][]{Lister:2001}.

Our detailed study of 3C~433 \citep[][]{Ramirez:2009} suggested that the near-IR polarisation of the core ($8.6\pm1$ per cent) is produced by  highly efficient dichroic extinction ($A_V\approx 15$ mag). Moreover, the polarised core in Cygnus~A \citep[$28\pm1$ per cent;][]{Tadhunter:2000a}, could also be explained in terms of in a similar fashion. In addition to the detection of highly polarised compact core sources, an interesting extended polarisation pattern has been detected along one edge of the bi-cone detected at 2.0~$\mu$m in Cygnus~A \citep{Tadhunter:2000a}, providing potentially valuable information on the anisotropic illumination by the AGN and the properties of the obscuring torus structure. The detected polarisation pattern in Cygnus~A \citep{Tadhunter:2000a} has been explained in terms of the illumination of the extended structures by anisotropic near-IR radiation emitted by the outer parts of a warped accretion disc \citep{Sanders:1989,Tadhunter:2000a}. The detection of similar patterns in other NLRG would support the idea that the near-IR continuum is intrinsically anisotropic. Such studies constitute a motivation to extend the polarimetric analysis to a larger sample of radio galaxies.

In this paper we present near-IR polarisation measurements for a sample of FRII NLRG taken with the {\em HST}. The sample and observations are described in section \ref{observations}, and the data reduction is presented in section \ref{Polarimetry}. In section \ref{Polarimetric_analysis} we analyse the polarimetric data,  and report the degree of intrinsic nuclear polarisation, as well as evidence for extended polarisation structures. We examine the possible origin of the core polarisation in section \ref{polarisation_mechanisms}, and finally, discussion and conclusions are given in sections \ref{Discussion} and \ref{Conclusion}, respectively.

\section[]{Sample and observations}\label{observations}

The sample comprises all 10 NLRG  at redshifts \mbox{$0.03<z<0.11$} in the 3CRR catalogue \citep*{Laing:1983} classified as FRII \citep[][]{Fanaroff:1974} sources based on their radio morphology; we label this sample as the `complete sample'. In addition, we have analysed archival observations of 3C~293, 3C~305 and 3C~405 (Cygnus~A) extracted from the Mikulski Archive for Space Telescopes (MAST). They are included because these are the only other 3C PRG at similar redshifts observed with {\em HST}/NICMOS at 2.05~$\mu$m in a similar way to our complete sample. These three additional objects, along with the 10 objects in the main sample, comprise an `extended sample' of 13 3CR radio galaxies, all located in the northern hemisphere. 

Dedicated {\em HST} polarimetric observations were obtained with the Near Infrared Camera and Multi-Object Spectrometer 2 (NICMOS2), through the three long polarimeter (POL-L) filters, centred at 2.05~$\mu$m. The principal axes of the three polarisers are offset by $\sim\!60^{\circ}$ from each other \citep{Viana:2009}. NICMOS2 has field of view of $19.2\times19.2$ arcsec. The observations of the complete sample were made during Cycle 13, between April 2005 and June 2006 (GO~10410, principal investigator, PI: C. N. Tadhunter). The observations were executed in multiple accumulate mode\footnote[2]{Multiple accumulate mode (multiaccum) is a pre-defined sequence of multiple non-destructive readouts exposures, used to cope with saturated pixels and cosmic rays, and to increase the dynamic range of the observations, i.e., increase the charge capacity of the pixels.}. 

The observations for GO~10410 were taken with the standard {\sc x-strip-dither-chop} pattern \citep{Viana:2009}. The spacing of the chop in the Y-direction was $31.5$ to $35.5$ arcsec, and the dither spacing in the X-direction was $0.9$ arcsec. A total on-galaxy integration time of $1024$~s (4 dithers of $256$ s each: $\sim$0.5 orbits in total) for each POL-L filters was achieved, with the exception of the relatively bright source 3C~433, for which shorter exposure times and more dithers were use in order to avoid saturation \citep[4 dithers of 16 s each, and 12 dithers of 64 s each, for each POL-L filters; more details in][]{Ramirez:2009}. Similar off-galaxy background exposures were used to facilitate accurate background subtraction without compromising the signal-to-noise ratio in the nuclear regions of the galaxies. 

For 3C~293 and 3C~305 we used archive data from the observation program GO~7853 (PI: N. Jackson). These data were taken on 1998 August 19, using NICMOS2 with the POL-L filters at 2.05~$\mu$m, and were made by executing a {\sc nic-spiral-dith} pattern with 2 points in steps of $3.5$ arcsec \citep{Viana:2009}, achieving an on-galaxy exposure time of $2176$ s per POL-L filter. 

For 3C~405, the observations were taken from program GO~7258 \citetext{PI: C. Tadhunter; detail in \citealp{Tadhunter:1999} and \citealp{Tadhunter:2000a}}. These observations  were executed on 1998 December 16 in the same fashion as the GO~10410 program: using coordinated parallel exposures observations as part of a chop process. The polarimetric observations with POL-L filter of 3C~405 were executed with a {\sc four-chop} pattern with chop steps of $32.5$ arcsec in $28$ exposures: 14 on the sky and 14 on the source, with a total on-source exposure time of $2688$ s in each of the POL-L filters (14 exposures of $192$ s each). 

The {\em HST} observational details are presented in Table \ref{tableobservat}. The {\em HST} provides an actual spatial resolution of  $0.22$ arcsec (and a theoretical spatial resolution of $0.21$ arcsec) at  2.05~$\mu$m. This is much better than achieved in routine ground-based observation without adaptive optics. Moreover, the point spread function (PSF) is more stable in space than it is from the ground -- an important advantage for polarimetric imaging. The low redshift of the objects in the sample, and the {\em HST} polarimetric capabilities, allow the measurement of the near-IR core polarisation at unprecedented resolution -- on the 150 to 850 parsecs scales depending on redshift\footnote[3]{Throughout this paper the following cosmological parameters are assumed: $H_0=70$ km s$^{-1}$ Mpc$^{-1}$, $\Omega_M=0.3$, and $\Omega_{\Lambda}=0.7$.}.

\begin{table}
\begin{center}
  \caption{{\em HST} observation details. The column POL-L NICMOS2 (NIC2) gives the exposure time for each of the three POL-L filters at different angles.}
  \begin{tabular}{@{}l@{}c@{}c@{}cc@{}}
\hline
  Source &   Obs. ID & $\;$Filter exp. time (s)  & Date & $\textrm{PA}\_V3$ \\
                     &             &\multicolumn{1}{c}{ POL-L NIC2}& &  ($^{\circ}$)\\
 \hline
  3C~33  & 10410 & 1024 & 2005-07-10 & 67.8\\ 
  3C~98   & 10410 &  1024 & 2005-08-28 & 79.2\\
  3C~192  & 10410 &  1024  & 2005-06-15 & 282.3 \\
  3C~236  & 10410 & 1024 & 2005-06-15 & 278.6\\
  3C~277.3  & 10410 &  1024 & 2005-08-03 & 272.8  \\
  3C~285   & 10410 &  1024  & 2005-11-21 & 143.6 \\
  3C~321   & 10410&  1024 & 2005-08-22 & 277.5 \\
  3C~433   & 10410& 832 & 2005-08-22 & 343.8\\
  3C~452   & 10410 & 1024 & 2006-06-15 & 64.4\\
  4C~73.08   & 10410 &  1024 & 2005-04-21 & 300.9\\
 \hline 
  3C~293  & 7853&  2176  & 1998-08-19 & 265.5 \\
  3C~305  & 7853& 2176  & 1998-07-19 & 292.1 \\
  3C~405  & 7258 & 2688 &1997-12-15&214.9 \\
 \hline
\end{tabular}\label{tableobservat}
\end{center}
\end{table}

 \section{Data reduction}\label{Polarimetry}
 
 All raw NICMOS/{\em HST} data were processed using the standard NICMOS pipeline calibration software {\sc calnica} \citep{Thatte:2009}. The {\sc calnica} pipeline removes instrumental signatures and cosmic ray hits, and combines the multiple readouts when observed in multiaccum mode. Subsequently, we further reduced the {\sc calnica} output with {\sc iraf} (Image Reduction and Analysis Facility)\footnote[4]{{\sc iraf} is distributed by the National Optical Astronomy Observatory, which is operated by the Association of Universities for Research in Astronomy, under cooperative agreement with the National Science Foundation.}, median combining the dithers on the source, and the dithers on the sky,  with {\sc imcombine}. The combined sky frame was then subtracted from the combined source frame. Possible remaining hot pixels and detector quadrant features were removed by hand using the {\sc clean} task in the {\sc figaro} package of the {\sc starlink} software library.

To obtain the measurements of polarisation degree ($P$) and E-vector $\textrm{PA}$ we applied the following steps.
 
\begin{itemize}
\item  The mean and standard deviation in units of counts per second of the intensity in each POL-L filter image were measured (whether measured in individual pixels or apertures). For those objects with just two dither images (namely 3C~293 and 3C~305), this approach is not statistically correct, and to estimate the standard deviation a combination of the detector read noise and the Poisson noise in the signal in each dither position was considered. 

\item The Stokes parameters, $I$, $U$ and $Q$, were solved from the matrix of the linear system for the three polarisers using a purpose-written {\sc interface description language} ({\sc idl}) routine that we wrote following the procedures of \citet{Sparks:1999}. We used the high-accuracy transmission coefficients for the POL-L filters determined for post-NICMOS cooling system observations by \citet{Batcheldor:2009} for objects observed after 31st December 1998, and the pre-NICMOS cooling system coefficients for the observations taken before then \citep*{Hines:2000}. 

\item $P$ and $\textrm{PA}$  were calculated  using the relationships $P=\sqrt{Q^{2} + U^{2}}/I$ and $\theta=\frac{1}{2}{\rm tan}^{-1}(U/Q)$. To get the $\textrm{PA}$, the aperture position angle [AP\textrm{PA} from the image headers (AP\textrm{PA}$=\textrm{PA}\_V3-405^{\circ}$)] had to be subtracted from $\theta$. The orientations of the NICMOS camera relative to the celestial frame  of the observations ($\textrm{PA}\_V3$), which must be subtracted from the calculated $\textrm{PA}$, are given in Table \ref{tableobservat}.
\end{itemize}

\section{Polarimetric analysis}\label{Polarimetric_analysis}

\subsection{Nuclear polarisation: aperture polarimetry}\label{aperture:polarimetry}

\begin{figure*}
\begin{tabular}{c@{}c@{}c@{}c}
\includegraphics[width=4.4cm,angle=0]{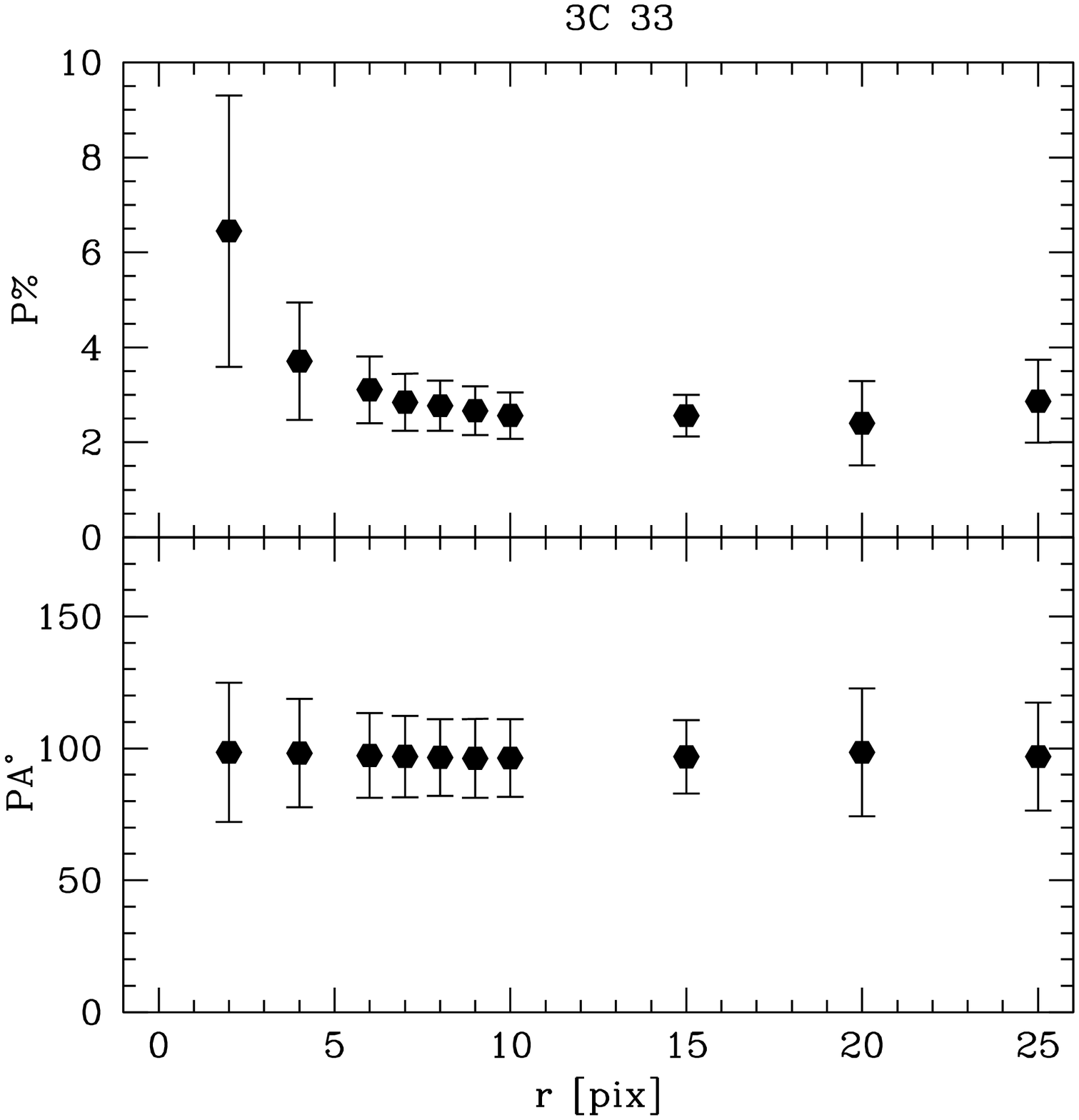}
&\includegraphics[width=4.4cm,angle=0]{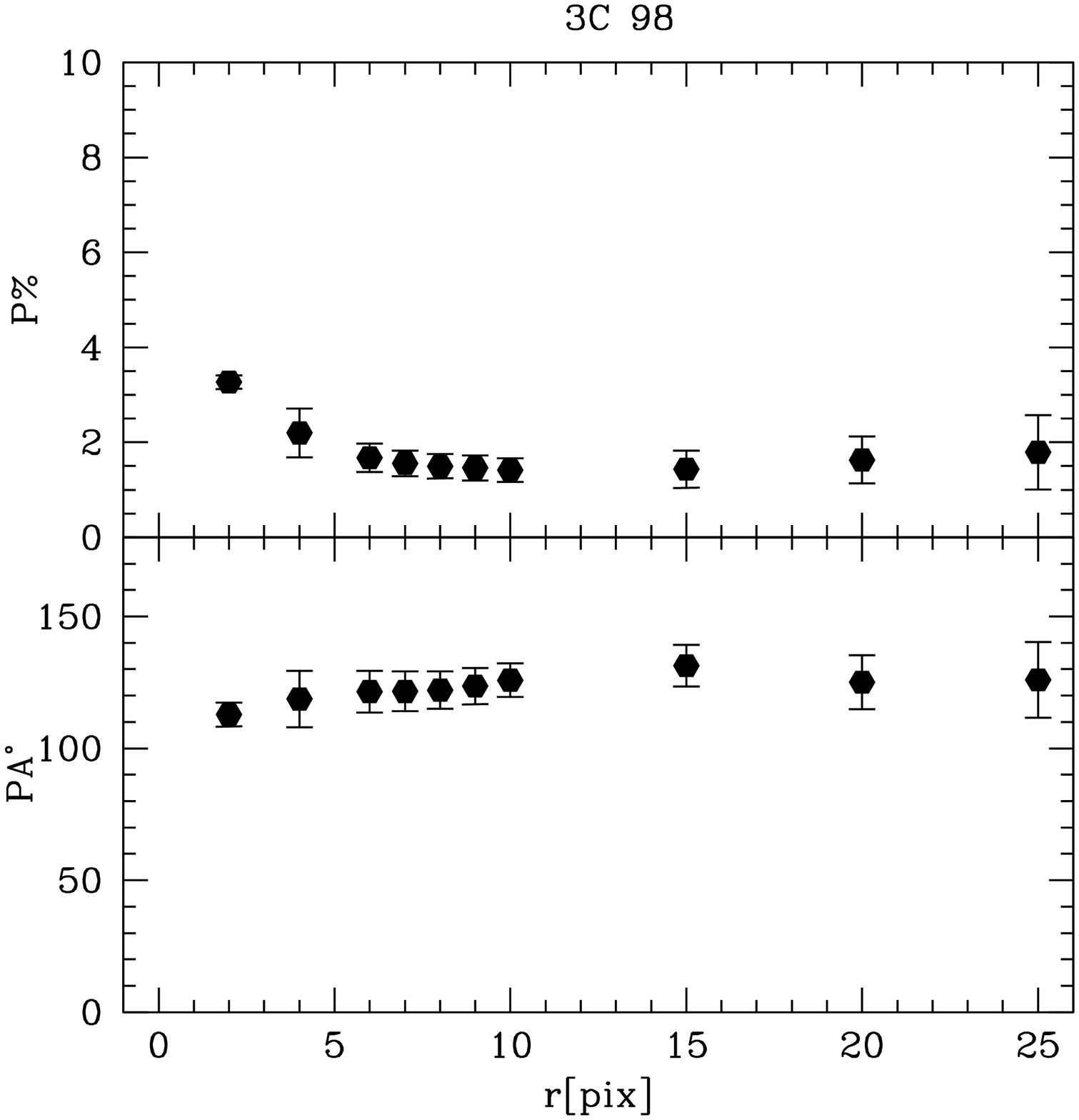}
&\includegraphics[width=4.4cm,angle=0]{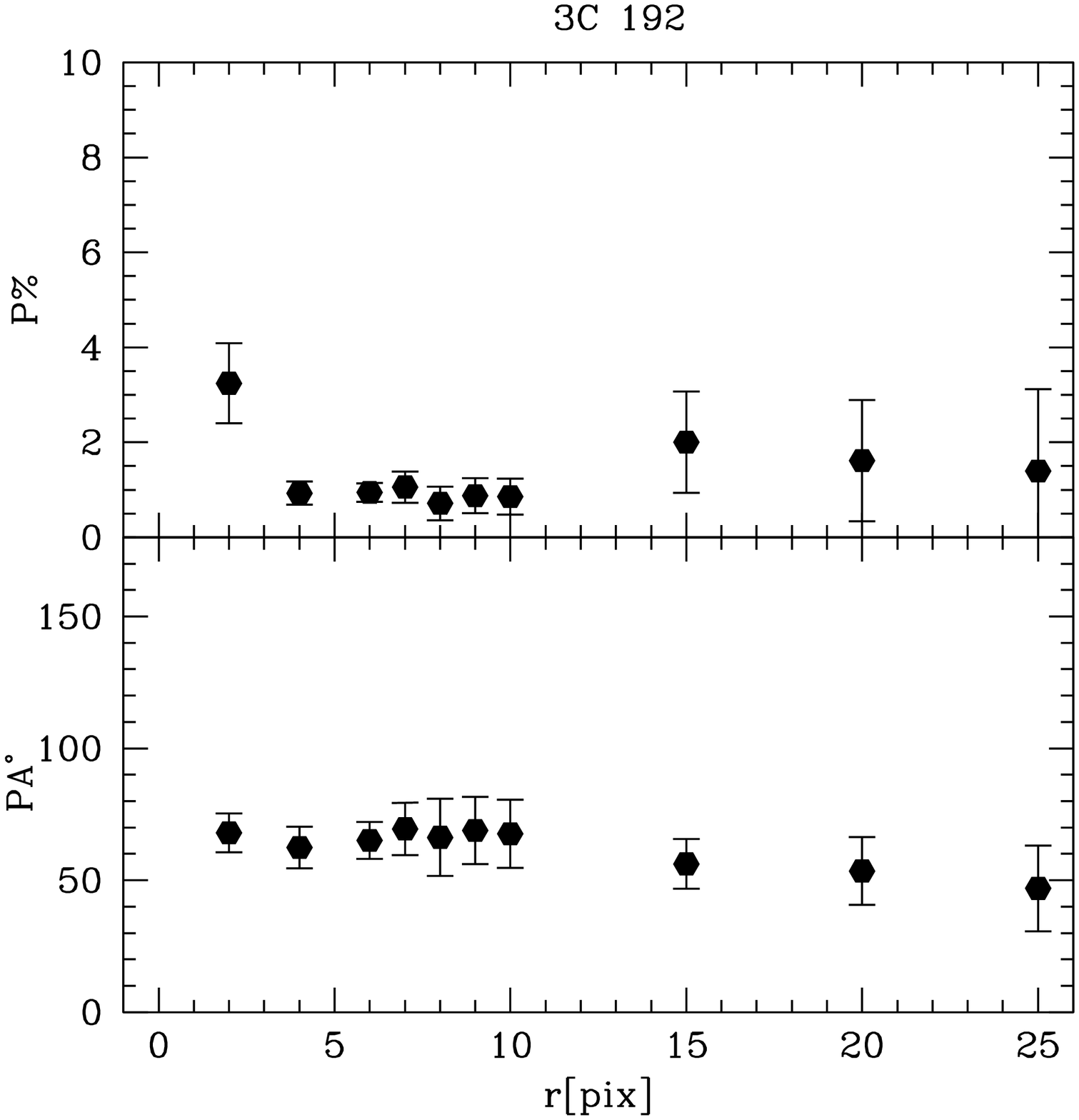}
&\includegraphics[width=4.4cm,angle=0]{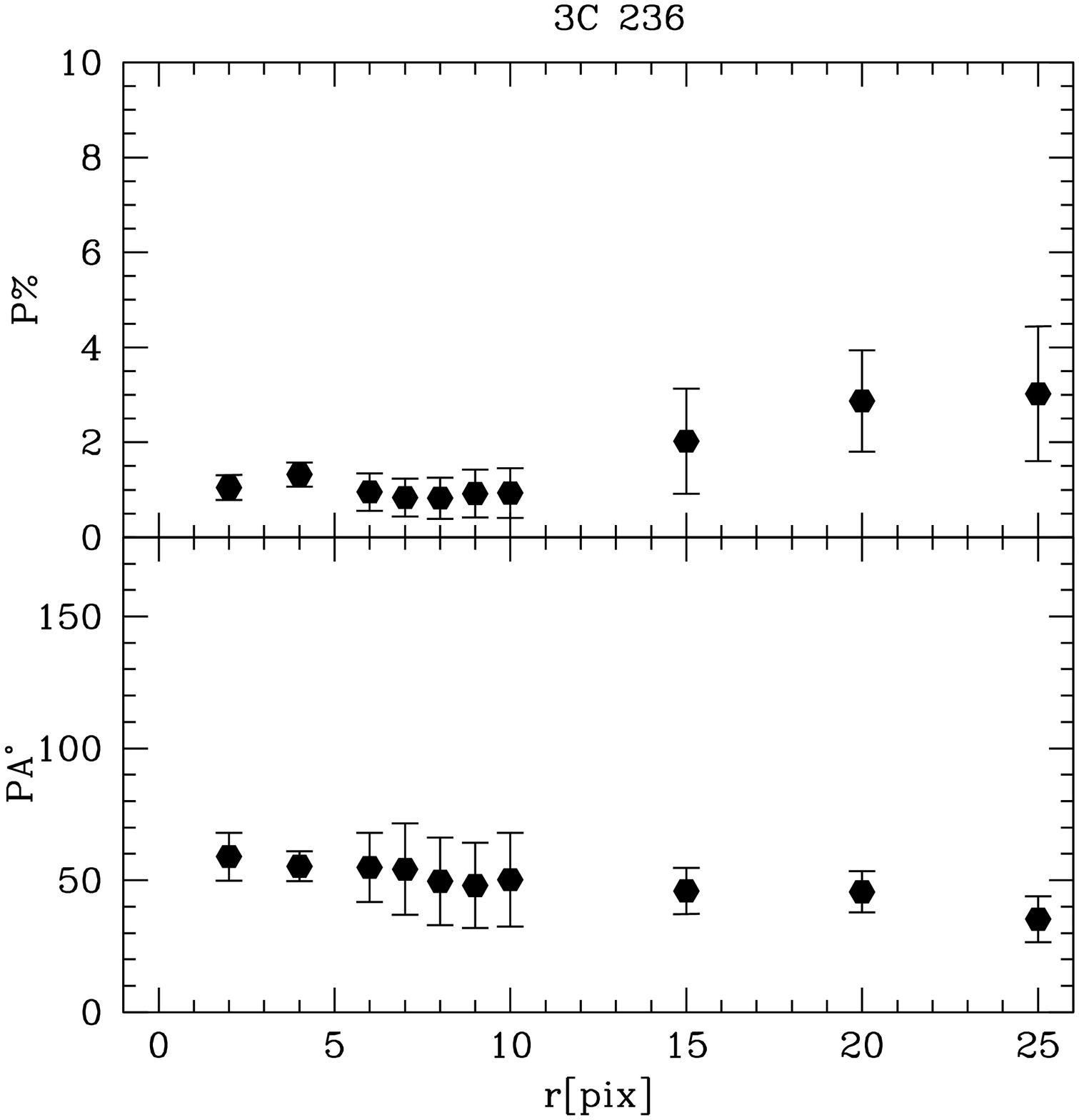}
\\
\includegraphics[width=4.4cm,angle=0]{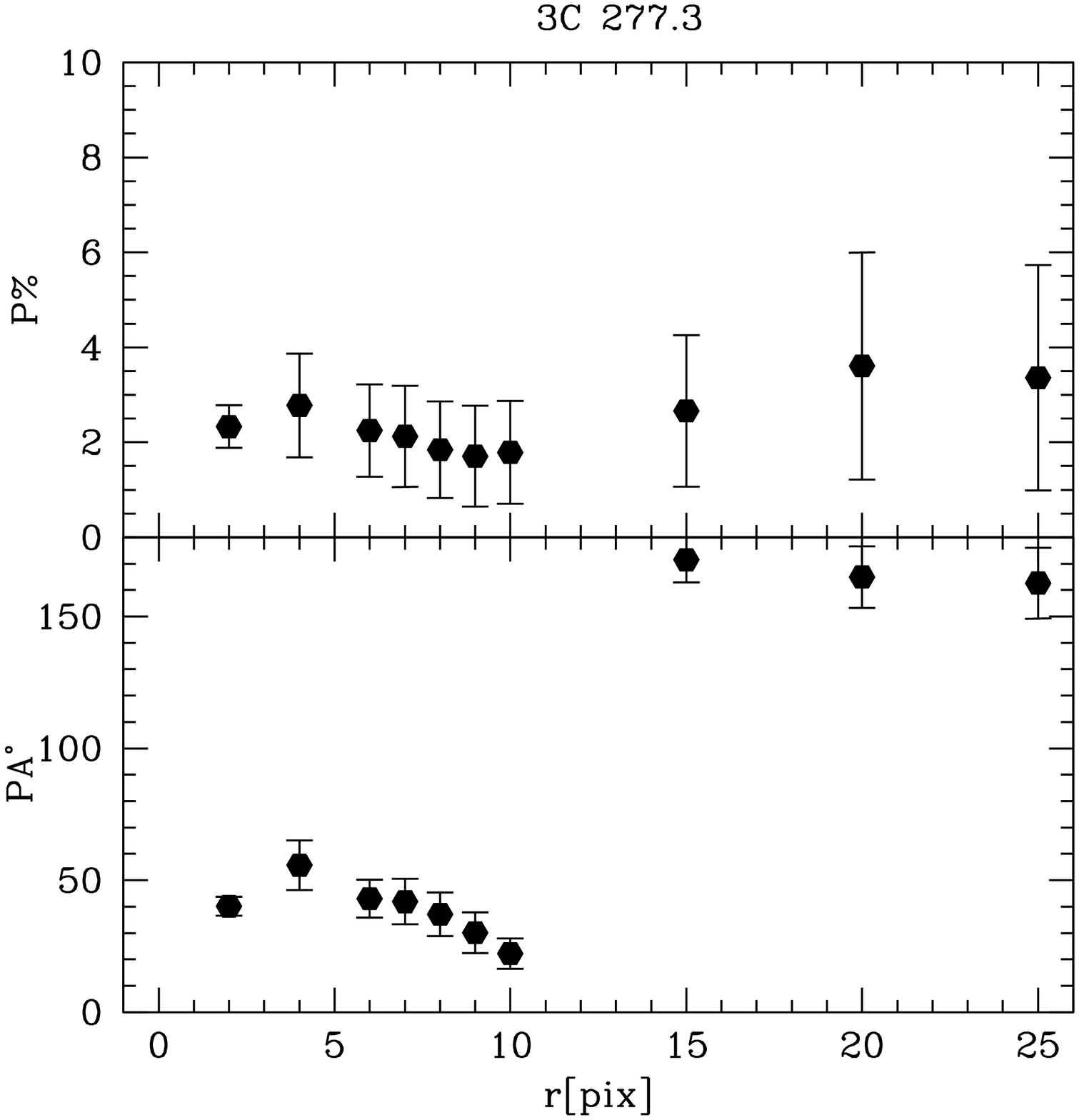}
&\includegraphics[width=4.4cm,angle=0]{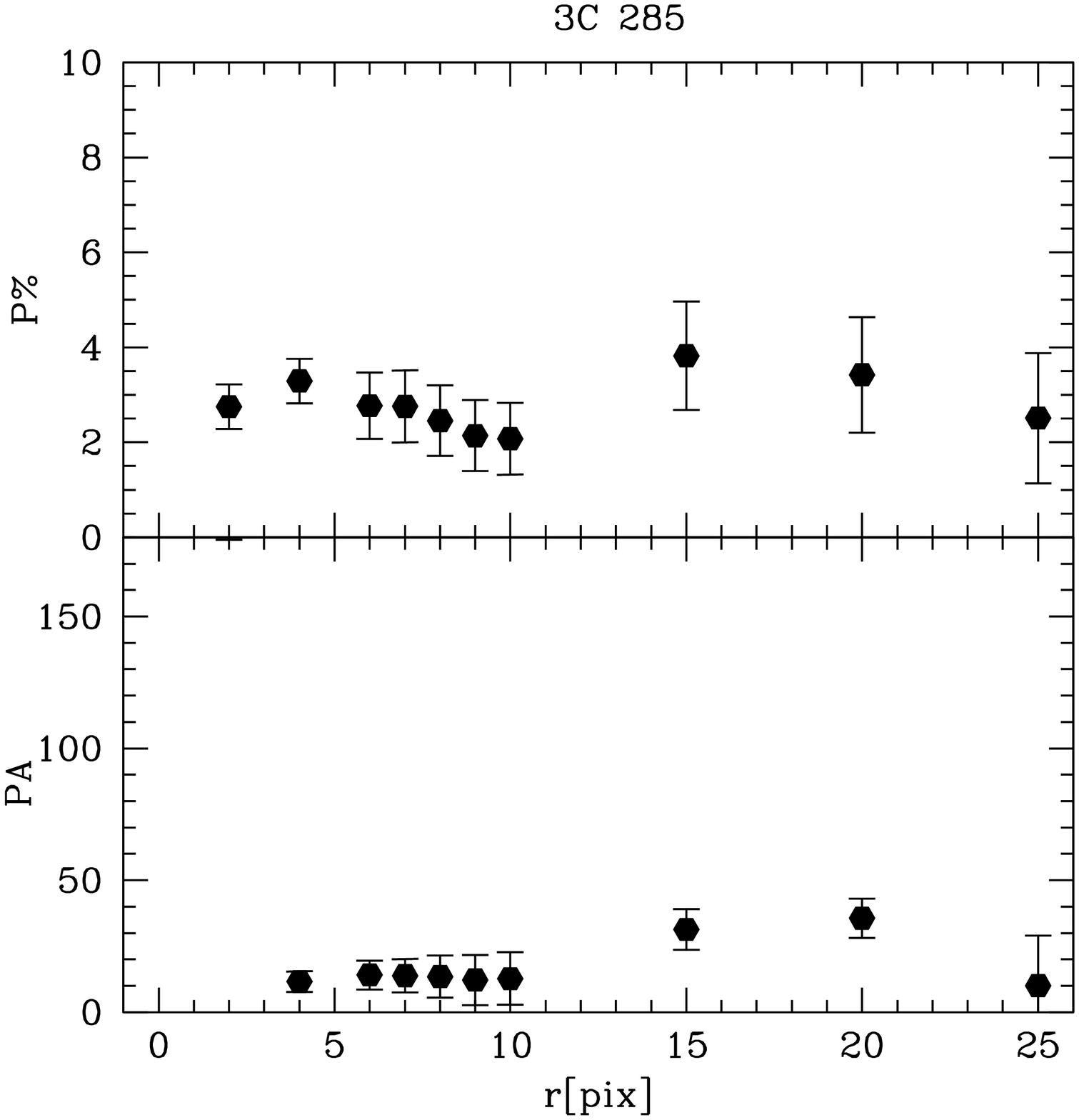}
&\includegraphics[width=4.4cm,angle=0]{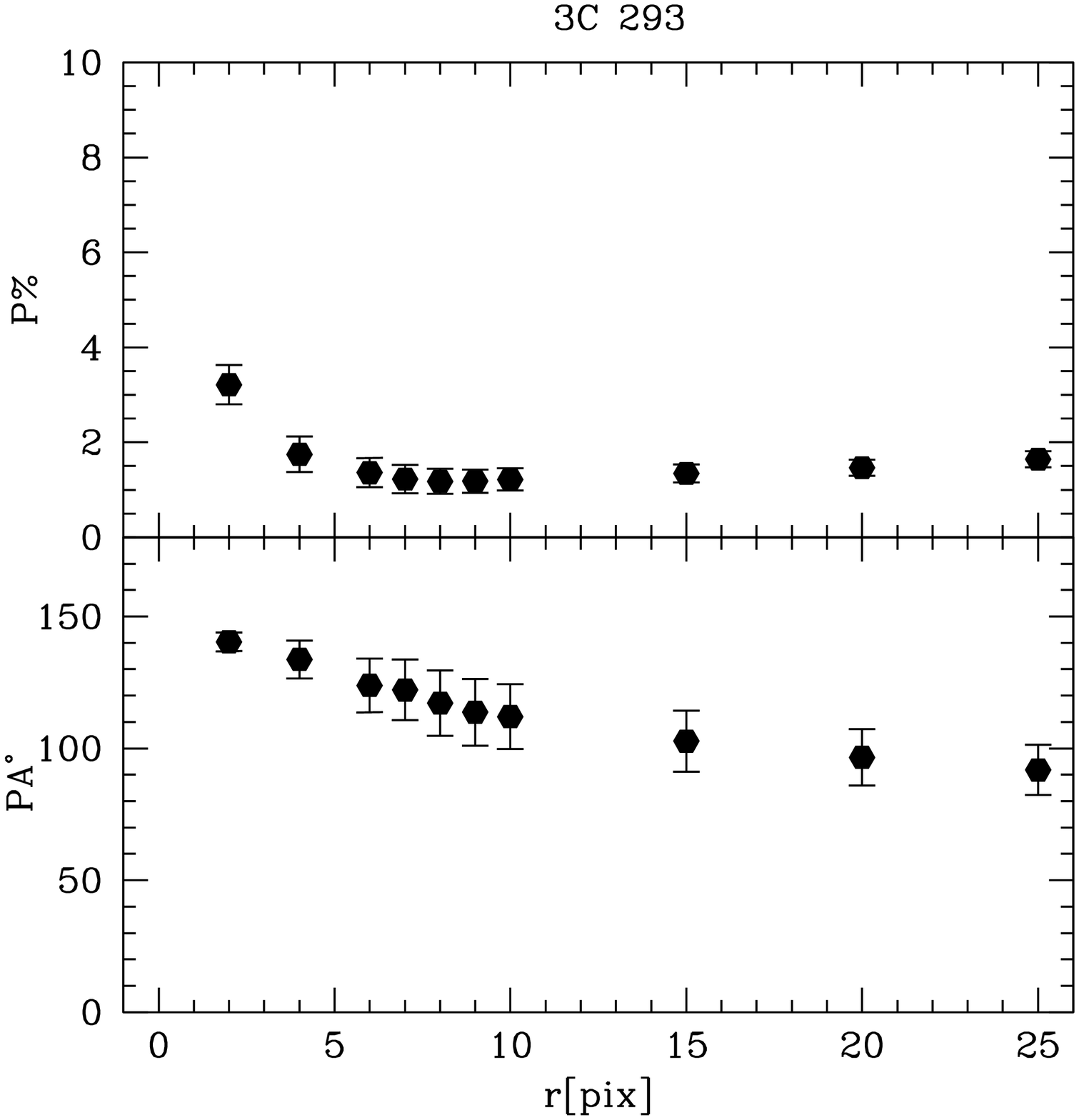}
&\includegraphics[width=4.4cm,angle=0]{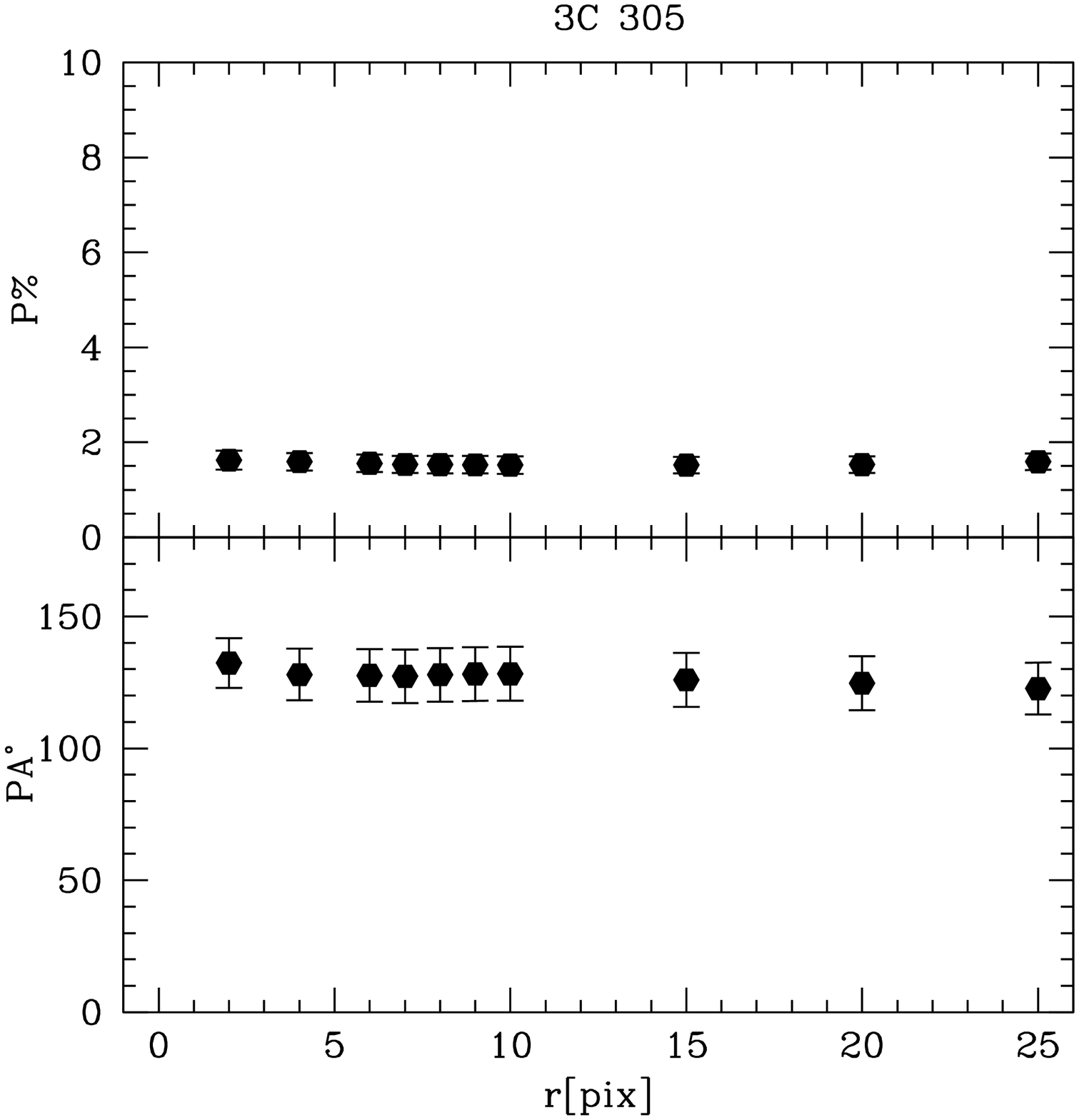}
\\
\includegraphics[width=4.4cm,angle=0]{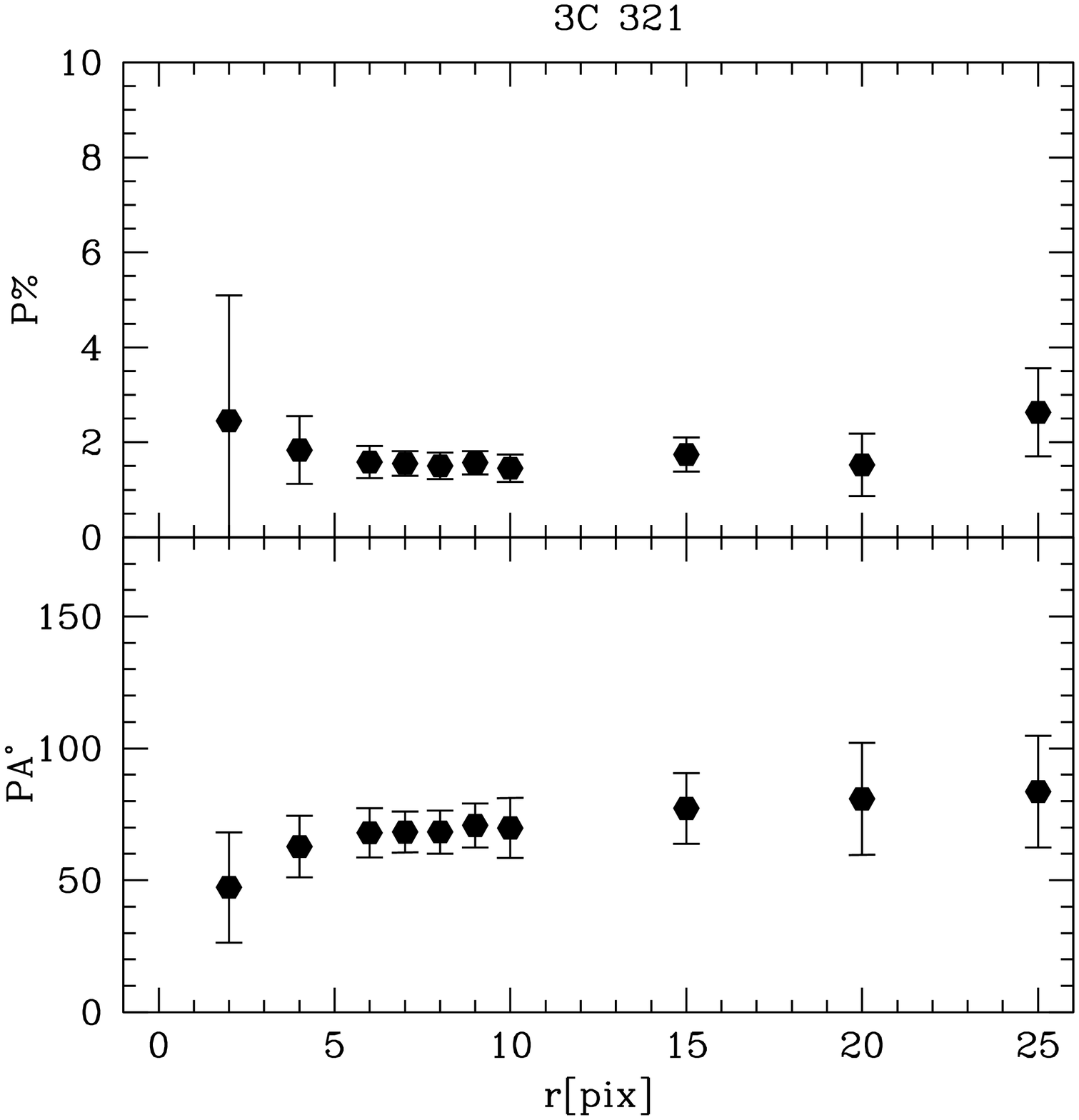}
&\includegraphics[width=4.4cm,angle=0]{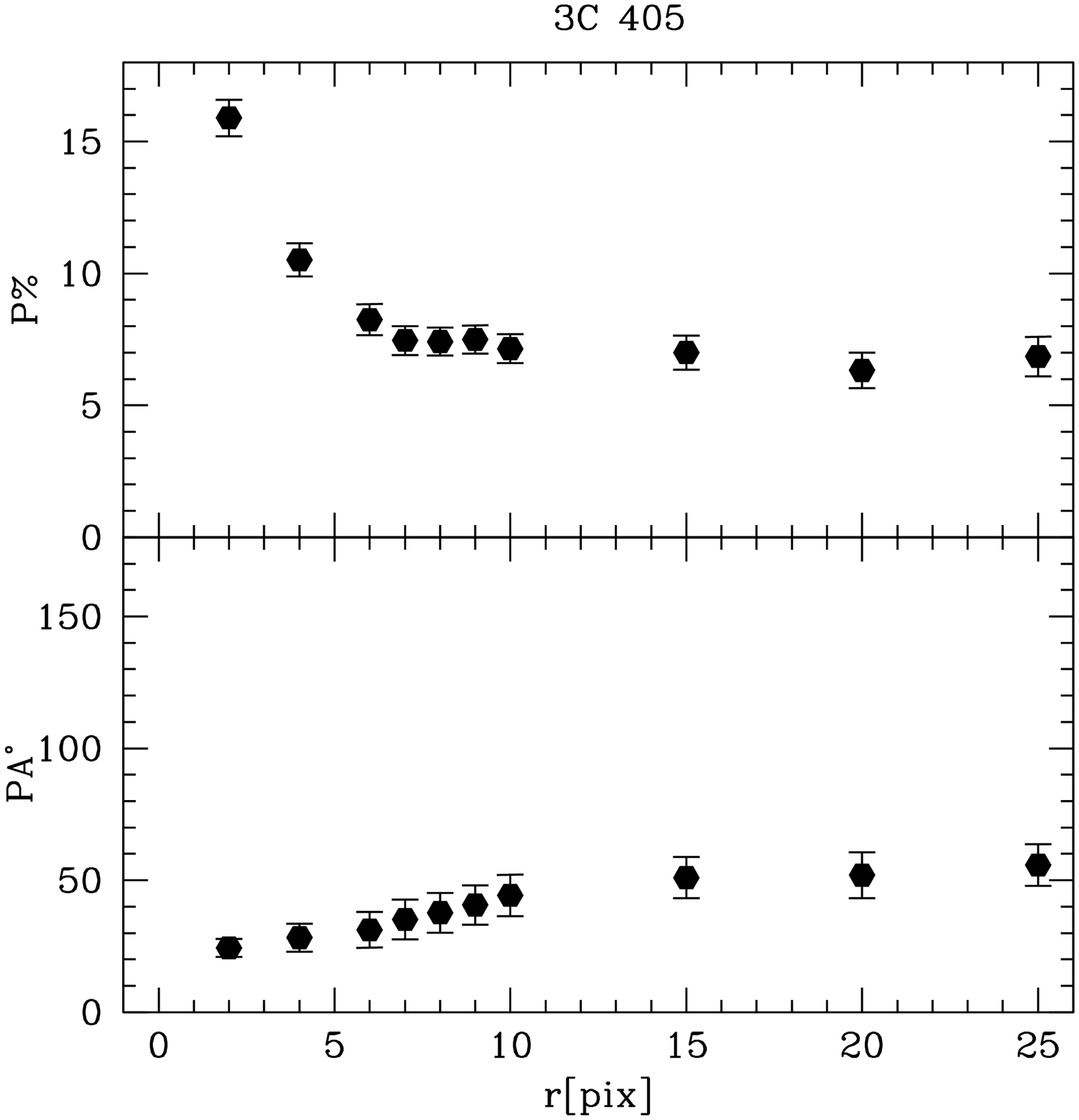}
&\includegraphics[width=4.4cm,angle=0]{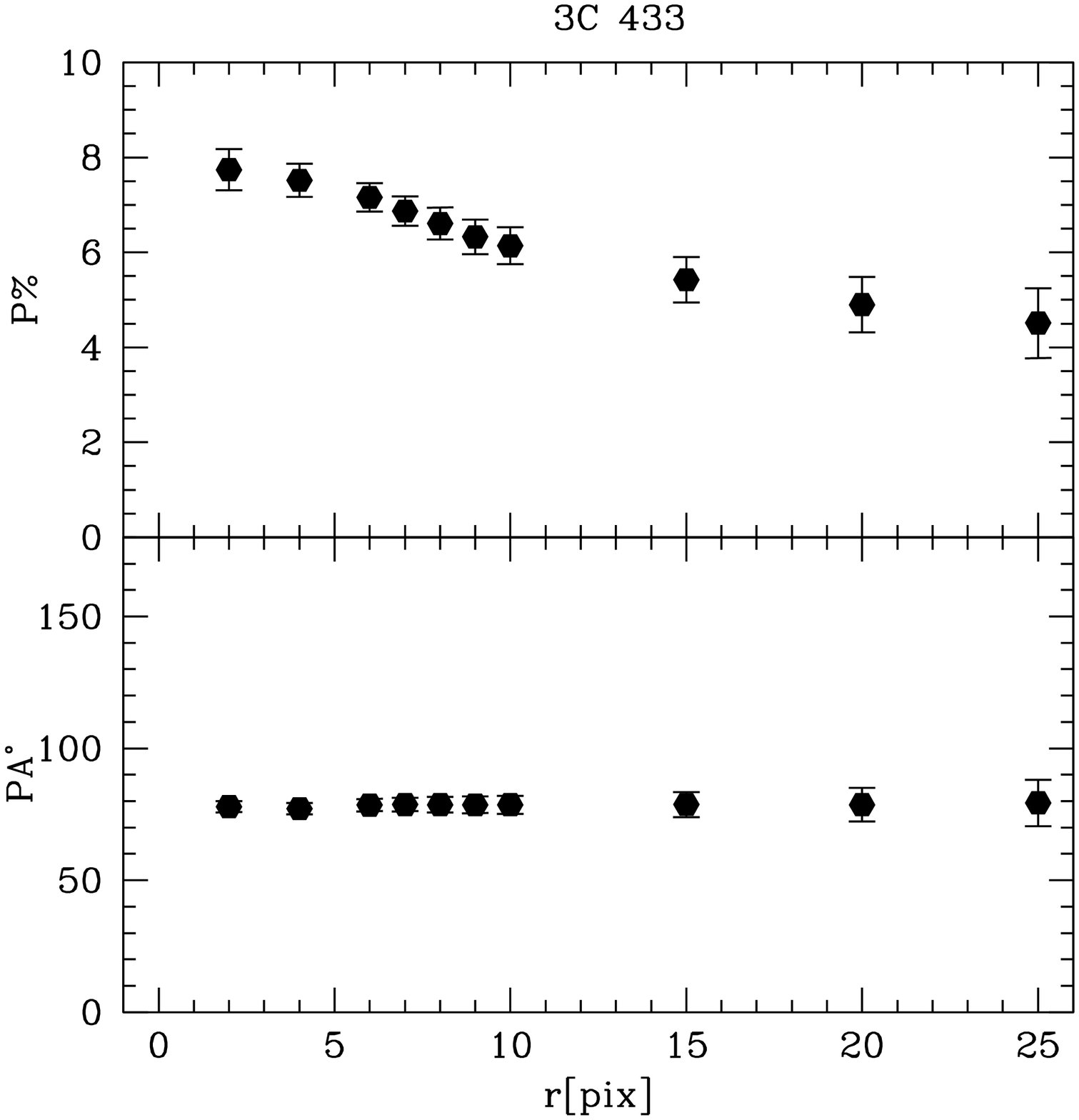}
&\includegraphics[width=4.4cm,angle=0]{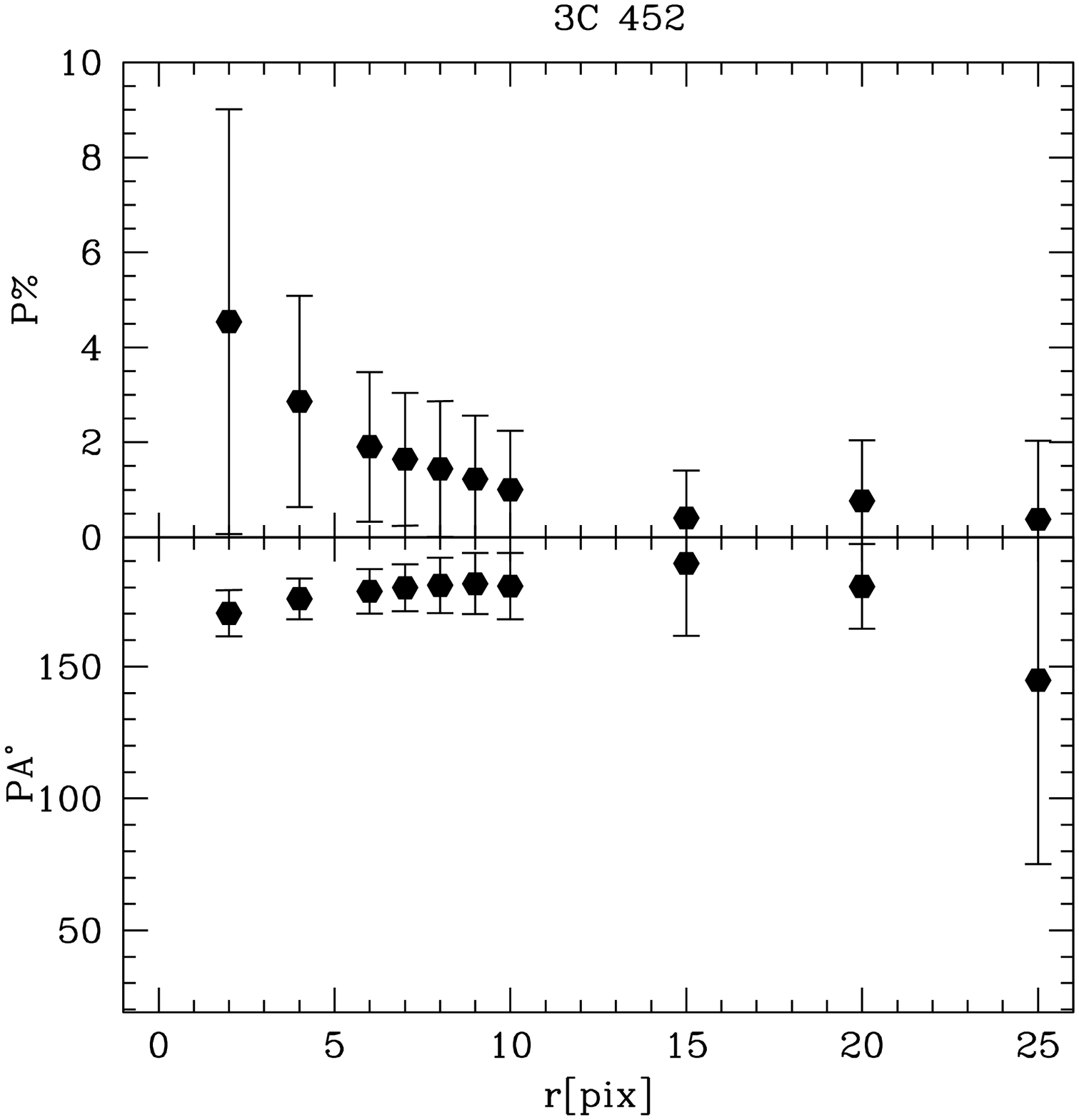}
\end{tabular}\\\raggedleft
\parbox{4.5cm}{
\includegraphics[width=4.5cm,angle=0]{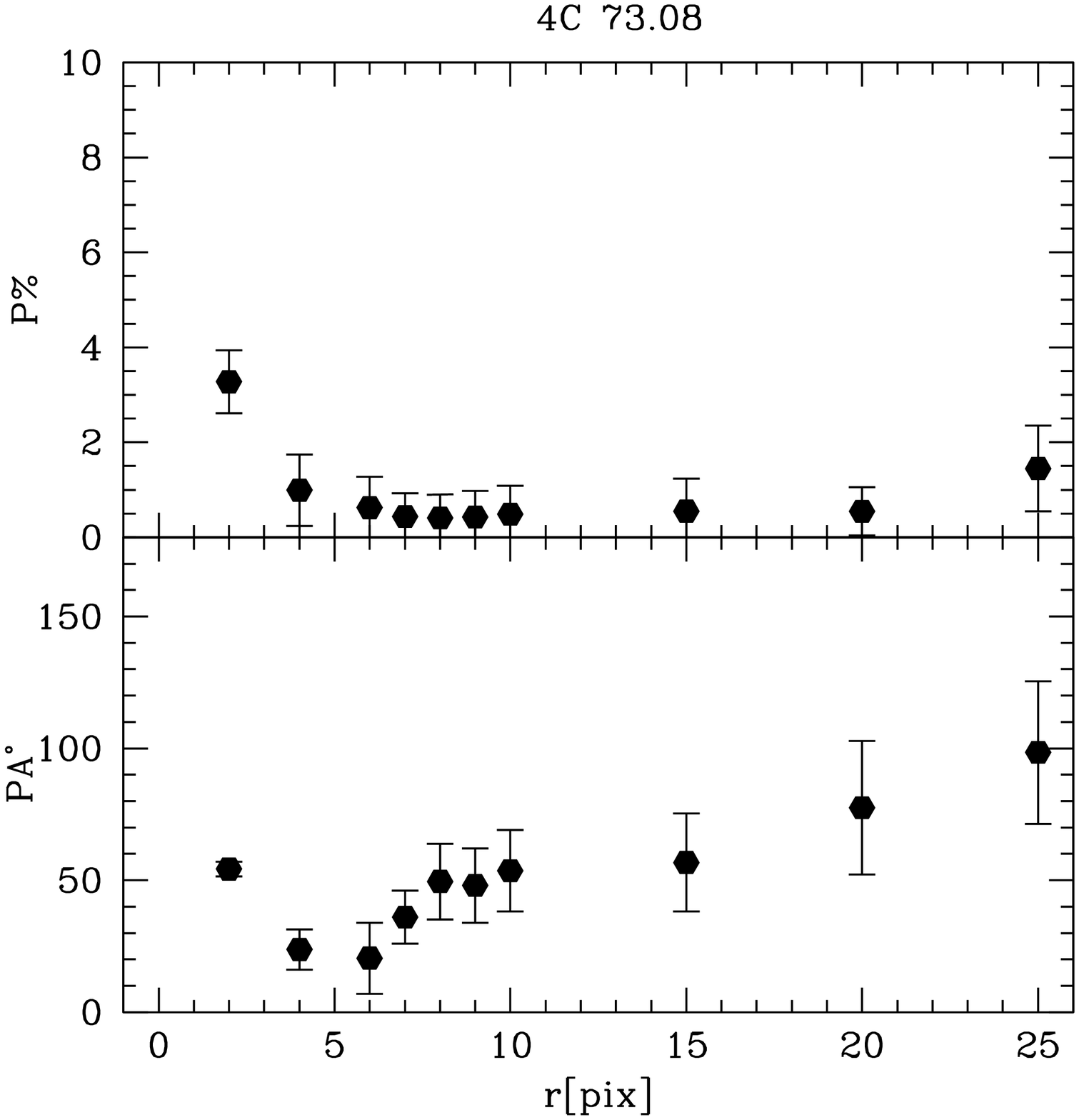}
}\hspace{1cm}\parbox[t]{12cm}{\caption{Polarisation measured at 2.05~$\mu$m through apertures of increasing radius. Polarisation ($P$) at the top, and position angle (PA) at the bottom.}\label{tablepolaperture}}
\end{figure*}

To measure the nuclear polarisation, aperture photometry of the galaxies was performed using several circular apertures of different radii centred on the nucleus using {\sc phot} task in {\sc iraf}. The central position of each galaxy was estimated by fitting a 2D Gaussian profile to the 2.05~$\mu$m deep image. The Stokes parameters were solved for each concentric aperture following the prescription of \citet{Sparks:1999}, as described in Section \ref{Polarimetry}. The variations of $P$ and $\textrm{PA}$ with aperture radius are presented for each source in Fig \ref{tablepolaperture}. In most cases the degree of polarisation shows a decrease with increasing aperture size, while the polarisation $\textrm{PA}$ shows relatively little variation. This is indicative of dilution of the core polarisation by unpolarised starlight toward the outer parts of the galaxies.

The core polarisation was considered to be that measured in an aperture of $0.9$ arcsec diameter (6 pixel radius), centred on the nucleus. This aperture size was chosen because it contains the first Airy ring (diameter of first Airy ring is $\sim\! 0.6$ arcsec), is not sensitive to small errors in the spatial alignment between data taken with the three POL-L filters, and does not contain too much of the light of the extended stellar halo of the host galaxy.

The measured polarisation estimates given by the relation $P=\!\sqrt{Q^{2} + U^{2}}/I$ are biased towards non-zero polarisation \citep{Clarke:1986,Sparks:1999}. Since $P=\!\sqrt{Q^{2} + U^{2}}/I$, negative values of $Q$ and $U$ will always give a positive length of the $P$ vector, resulting in a residual observed polarisation \citep{Clarke:1986}. This bias affects targets with low signal-to-noise ratio, $P/\sigma\lesssim3$, resulting in the measurement of polarisation even when the source is truly unpolarised.  The distribution function for the polarisation, $P$, and the true polarisation, $P_0$, obeys a distribution governed by a Ricean distribution \citep{Rice:1945}:
\begin{equation}\label{Ricean}
F(P,P_0)=P e^{-[(P^2+P_0^2)/2]} I_0(PP_0)
\end{equation}
where $ I_0(PP_0)$ is the zero-order Bessel function \citep{Clarke:1986,Sparks:1999}. On the other hand, the instrumental polarisation of NICMOS2 at 2.05~$\mu$m, caused by the optical path in the telescope and NICMOS, gives an upper limit of $0.6\pm0.1$ per cent which affects the polarisation of targets with polarisation lower than $\sim\! 1$ per cent \citep{Batcheldor:2009}. Overall, a good polarisation detection requires $P/\sigma > 3$ and $P>0.6$ per cent.

Polarisation has been detected with $P/\sigma>3$ confidence in 9 out of the 13 sources. For 3C~236 and 3C~277.3 ($P^{\rm obs}=1.0\pm0.4$ and $P^{\rm obs}=2.1\pm0.9$, respectively) the detected polarisation is marginal, with $2\!< P/\sigma<\!3$. Therefore, the results for these two low polarisation sources must be taken with caution. For 3C~452 and 4C~73.08 ($P^{\rm obs}=1.9\pm1.6$ and $P^{\rm obs}=0.6\pm0.7$, respectively) we find $P/\sigma<2$ -- too low to be considered significant detections of near-IR polarisation. We have corrected these four sources using the mean Ricean bias (see Table \ref{tablepolarisation}). However, 3C~452 and 4C~73.08 sources are not considered in the remaining analysis. Moreover, the measured polarisation of 4C~73.08 is the only one with $\leq\!0.6$ per cent -- the estimated level of instrumental polarisation -- another reason for this source not be considered further. Therefore, significant $P>\!2\sigma$  polarisation is detected in 8 out of 10 objects in the complete {\em HST} sample, and 11 out of 13 objects in the extended sample.

\begin{table}
  \caption{Near-IR polarisations of the nuclei of nearby 3CR FRII radio sources.
   $P$: polarisations measured through a $0.9$ arcsec diameter (6 pixels radius) apertures corrected for Ricean bias.
   $P^{\rm int}$: intrinsic nuclear polarisations of the unresolved point sources (assuming starlight unpolarised).
   $\textrm{PA}_{\rm E}^{\circ}$: E-vector position angle.
   $\textrm{PA}_{jet}^{\circ}$ position angle of the radio jet.
   $\Delta \textrm{PA}_{\rm E-jet}^{\circ}$: offset between the E-vector and radio jet.
  The `$\bot$' indicates perpendicularity within $\pm 20^{\circ}$.
   }\label{tablepolarisation}
  \begin{tabular}{@{}l @{ } l@{ } l@{ } l@{ } l@{  } l@{}}
  \hline
   Source & \multicolumn{1}{c}{$P\%$} & \multicolumn{1}{c}{$P^{\rm int}\%$ }& \multicolumn{1}{c}{$\textrm{PA}_{\rm E}^{\circ}$} & \multicolumn{1}{c}{$\textrm{PA}_{\rm jet}^{\circ}$} &$\;\Delta \textrm{PA}_{\rm E-jet}^{\circ}$\\
 \hline
  3C~33  & $3.1\pm0.7$  & $17.6\pm4.0$ & $97.3\pm16.1$  & $19\pm4$   & $ 78\pm17$$\bot$ \\ 
  3C~98  & $1.7\pm0.3$  & $60.5\pm11.2$ & $121.5\pm7.9$  & $23\pm3$   & $82\pm8$$\bot$ \\
  3C~192 & $0.9\pm0.2$ & $>54.0$ & $65.1\pm7.0$   & $120\pm5$  & $55\pm9$    \\
  3C~236 & $0.9\pm0.4^a$ & $5.6\pm2.5$  & $55.1\pm13.1$ & $125\pm5$  & $70\pm14$$\bot$\\
  3C~277.3& $2.0\pm0.9^a$ & $36.0\pm16.6$& $43.0\pm7.2$ & $156\pm5$  & $67\pm9$  \\
  3C~285 & $2.7\pm0.7$  & $16.2\pm4.2$ & $14.1\pm5.5$   & $80\pm3$   & $66\pm6$  \\
  3C~321 & $1.6\pm0.3$  & $>32.8$ & $68.0\pm9.3$  & $123\pm6$  & $55\pm11$ \\
  3C~433 & $7.0\pm0.2$  & $9.8\pm0.8$  & $83.0\pm2.3$   & $168\pm2$  & $85\pm3$$\bot$ \\
  3C~452 & $1.4\pm1.6^b$  & $6.6\pm7.6$& $178.5\pm8.5$ & $81\pm2$  & \multicolumn{1}{c}{---} \\
  4C~73.08& $0.4\pm0.7^b$ & $5.9\pm9.6$&$20.5\pm13.5$ & $70\pm3$   & \multicolumn{1}{c}{---} \\ 
\hline
  3C~293 & $1.4\pm0.3$ & $43.5\pm15.8$& $123.9\pm10.2$ & $95\pm5$  & $29\pm11$  \\
  3C~305 & $1.5\pm0.2$ & $31.5\pm5.2$ & $127.6\pm10.0$ & $55\pm5$     & $ 73\pm11$$\bot$  \\
  3C~405 & $8.3\pm0.6$ & $>34.7$ & $31.2\pm6.7$  & $105\pm2$   & $ 74\pm7$$\bot$\\
\hline
\end{tabular}
{\footnotesize {\bf Notes:} $^a\,2<P/\sigma<3$. $^b\,P/\sigma<2$. These have been corrected for Ricean bias.}
\end{table}

The intrinsic degree of the core polarisation is underestimated by these initial measurement because unpolarised starlight dilutes the intrinsic polarisation (under the assumption that the starlight is unpolarised). Also, the relatively large aperture may lead to some geometrical dilution of the polarisation due to the combination of polarisations with different $\textrm{PA}$, leading to lower net polarisation. To estimate the contribution of starlight relative to AGN light in the nuclear aperture ($I_{\rm stars}/I_{\rm AGN}$), we measured the total flux enclosed in a $0.9$ arcsec diameter aperture for the host galaxy after PSF subtraction \citep[$I_{\rm stars}$, see][ for details about the PSF subtraction]{Ramirez:2014}, and the PSF flux within same size aperture ($I_{\rm AGN}$). Using this fraction we estimated the intrinsic polarisation, $P^{\rm int}=P (1+I_{\rm stars}/I_{\rm AGN})$. The $\textrm{PA}$ of the E-vector is unaffected by the starlight.

The measured and intrinsic core polarisation and $\textrm{PA}$ in the $0.9$ arcsec diameter aperture are presented in Table \ref{tablepolarisation}. The uncertainty in the intrinsic polarisation is estimated from the uncertainty in the measured polarisation, as well as the estimated uncertainty in the fractional contribution of starlight in $0.9$ arcsec aperture. The measured polarisation ranges between $1$ and $9$ per cent. After applying starlight correction as described above, the intrinsic polarisation ranges between $6$ and $60$ per cent. This substantial increase following correction from starlight is due to the strong contribution to the flux from the underlying background galaxy hosting the AGN. Typically the AGN contributes only $< 20$ per cent in the nuclear aperture \citep[except 3C~433 with a contribution of $\sim70$ per cent; see][]{Ramirez:2014}. Note that, for cases in which we lack a significant detection of a nuclear point source at 2.05~$\mu$m \citep[namely 3C~192, 3C~321 and 3C~405:][]{Ramirez:2014}, we can only derive lower limits for the near-IR polarisations of their nuclei.

Regarding the $\textrm{PA}$ of the E-vectors, it has been found that, for 4 out of 8 (50 per cent) of the sources in our complete {\em HST} sample with significant near-IR polarisation, the polarisation E-vector is perpendicular to the radio jet axis within $\pm20\degree$, while in 6 out of 11 (54 per cent) sources in the extended sample, the E-vector is close to perpendicular to its radio jet axis (see Table \ref{tablepolarisation}).  For the rest of the sources, the E-vector is not perpendicular to the radio axis: their offset ranges between $29\degree$ and $67\degree$.  

We have used Kolmogorov-Smirnov test \citep[KS-test;][]{Kolmogorov:1933} to test the null hypothesis that the distribution of the position angle differences between the E-vectors and the radio jet axes ($\Delta \textrm{PA}_{\rm E-jet}^{\circ}$) is, in fact, drawn from a uniform distribution of $\textrm{PA}$ differences. We find that we can reject this null hypothesis at a 99 per cent level of significance, i.e., there is a 1 per cent chance that the distribution of $\Delta \textrm{PA}_{\rm E-jet}^{\circ}$ is drawn from a uniform distribution.

\subsection{Spatially-resolved polarimetry}\label{Extended:polarimetry}

Near-IR imaging polarimetry has the potential to penetrate the circumnuclear dust and trace structures on a kpc-scale. To investigate possible extended polarisation structures, a software package is available through the Space Telescope Science Institute (STScI) that produces polarisation coefficient images ($I$, $U$ and $Q$) from images taken with NICMOS \citep{Mazzuca:1999}. Nevertheless, we have developed our own {\sc idl} routine to allow a better manipulation of the images. This  {\sc idl} routine also executes binning of the images, solves the Stokes parameters, estimates $P$ and $\textrm{PA}$ (with their respective errors), and maps the $P$-vectors over the intensity image.

Because the spatial resolution of {\em HST} is $0.22$ arcsec at 2.05~$\mu$m (equivalent to $\sim\!3$ pixels sampling in NICMOS2), it is difficult to distinguish polarisation structures within a $3\times3$ pixels box. Therefore, to be sure of the degree of extended polarisation, each POL-L filter image was binned using a $3\times3$ pixel box. The value given to each pixel in the new low resolution image was set to be the sum of the original 9 pixels in the $3\times3$ pixels box. After the binning process, the Stokes parameters were solved for each new pixel of the lower resolution image.

\begin{figure*}
\includegraphics[width=78mm]{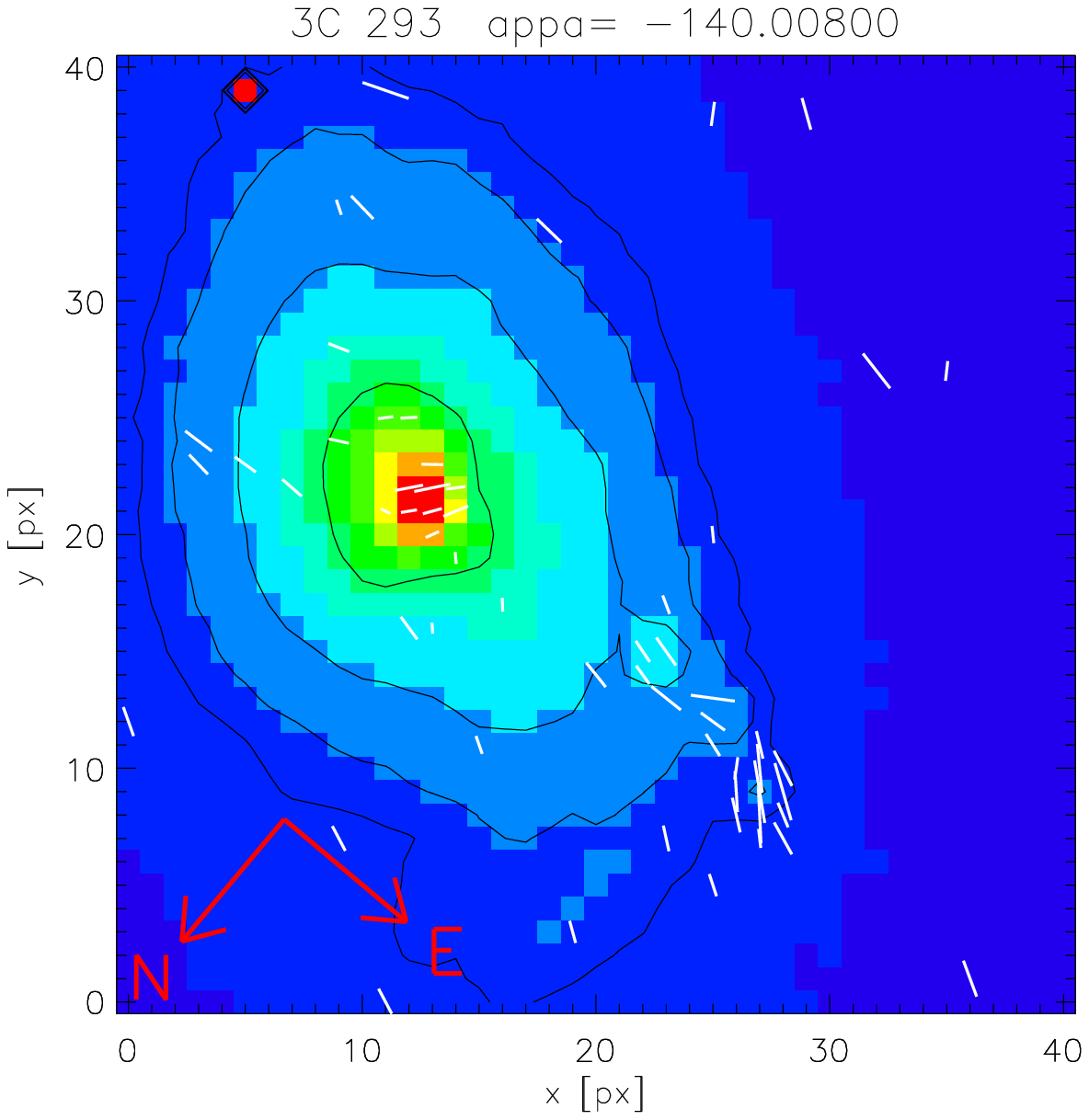}
\hspace{1em}
\includegraphics[width=78mm]{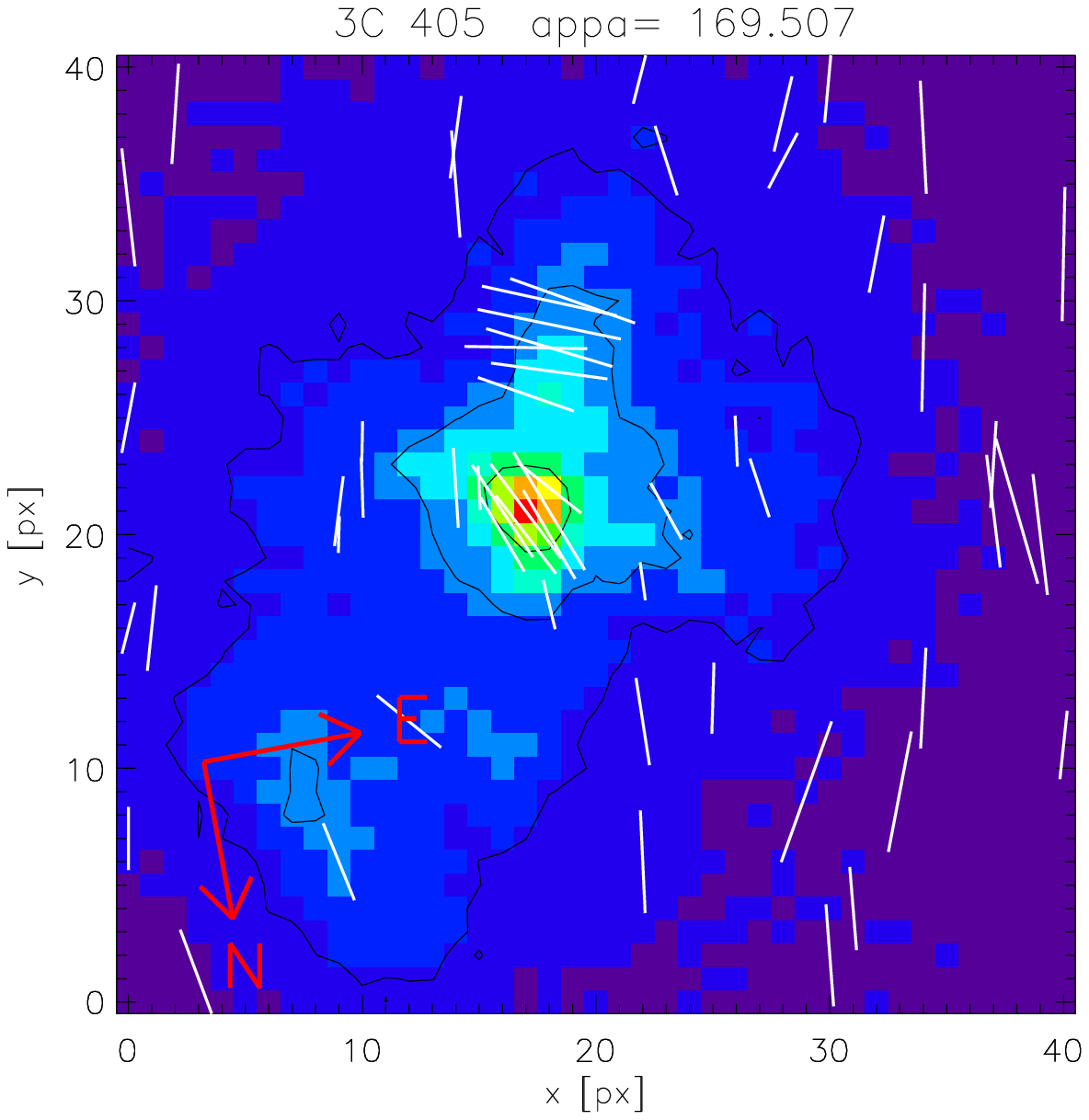}
 \caption{The 2.05~$\mu$m {\em HST} polarisation maps (E-vector) for 3C~293 (left) and 3C~405 (right, Cygnus~A), produced  with our in-house {\sc idl} routine. The  maps are consistent with those of \citet{Tadhunter:2000a} for Cygnus~A and \citet{Floyd:2006} for 3C~293, which were obtained using different software. The red arrows indicate north and east. The length of the longest polarisation vector represents $30\%$, and the polarisation vectors are only plotted for $P/\sigma>5$. In each case the field size is $3\times3$ arcsec.}
\label{polmaps}
\end{figure*}

Apart from the jet-knots detected in 3C~293 by \citet{Floyd:2006}, and the extended polarisation detected in 3C~405 by \citet{Tadhunter:2000a}, we have not found clear evidence for extended polarisation in any of the sources in the complete or extended sample. Figure \ref{polmaps} shows  our polarisation maps for 3C~293 and 3C~405. For confirmation of the correct operation of this in-house developed software, we compared the polarisation pattern to \citet{Tadhunter:2000a} for Cygnus~A, which were obtained using different software.

For the case of 3C~293, the near-IR jet comprises three knots running from west to east, following the radio jet counterparts \citep*{Bridle:1981}. We  measured slightly lower polarisation than the reported by \citet{Floyd:2006} for the three knots: whereas \citet{Floyd:2006} measured $15$ per cent polarisation for the three knots,  we measured  $12$ per cent for the farthest knot and $\approx 5$ per cent for the other two (see below). Following the nomenclature for the knots in \citet{Floyd:2006} ($E1$ is the farthest knot from the core, $E2$ is at intermediate distance, and $E3$ is the nearest one), we obtained the following results:
\begin{center}
\begin{tabular}{l@{ }l@{ }l}
$P_{E1}=12.6\pm1.3$ &per cent  and & $\textrm{PA}_{E1}=24.2\pm2.6^{\circ}$\\

$P_{E2}=5.0\pm0.5$ &per cent and &$\textrm{PA}_{E2}=84.8\pm7.9^{\circ}$\\

$P_{E3}=4.7\pm0.5$ &per cent and &$\textrm{PA}_{E3}=70.0\pm5.4^{\circ}$\\
\end{tabular}
\end{center}
Furthermore, we found an offset of $\sim\! 30$ degrees between our $\textrm{PA}_{E3}$ and that reported by \citet[][$\textrm{PA}_{E3}=40\degree$]{Floyd:2006}, while the other two knots are consistent, with an offset $<\!15 \degree$ \citep[$\textrm{PA}_{E1}=10\pm10^{\circ}$ and $\textrm{PA}_{E2}=70\pm10^{\circ}$;][]{Floyd:2006}. However, our measurements of the polarisation $\textrm{PA}$s of jet knots are similar to the polarisation angles measured at shorter radio wavelengths, where the Faraday rotation effects are minimised  \citep{Akujor:1996}. \citet{Akujor:1996} present 22 and 15 GHz polarimetric maps observed with VLA of the core and jet of 3C~293. Unfortunately, \citet{Akujor:1996} do not give numbers for the $\textrm{PA}$, but report that at high frequencies (22 GHz), the E-vectors run parallel to the jet \citep{Akujor:1996}. By examining the jet $\textrm{PA}$ in their 22 GHz image, we found consistency with our results.

\section{Examination of the polarisation mechanisms}\label{polarisation_mechanisms}

The polarisation of the near-IR light could be produced by three mechanisms: (i) dichroic extinction; (ii) non-thermal synchrotron radiation; or (iii) scattering of the light by dust or electrons in the near-nuclear regions. In this section we analyse each case critically.

\subsection{Dichroic extinction}
It is possible that the sources with a point source detection are dominated by dichroism, given that in such sources we may be observing the sources shining directly through the foreground dust in the torus \citep[point sources are detected at 2.05~$\mu$m in 3C~33, 3C~98, 3C~236, 3C~277.3, 3C~285, 3C~293, 3C~305, 3C~433, 3C~452 and 4C~73.08;][]{Ramirez:2014}.

\begin{figure}
\includegraphics[width=84mm]{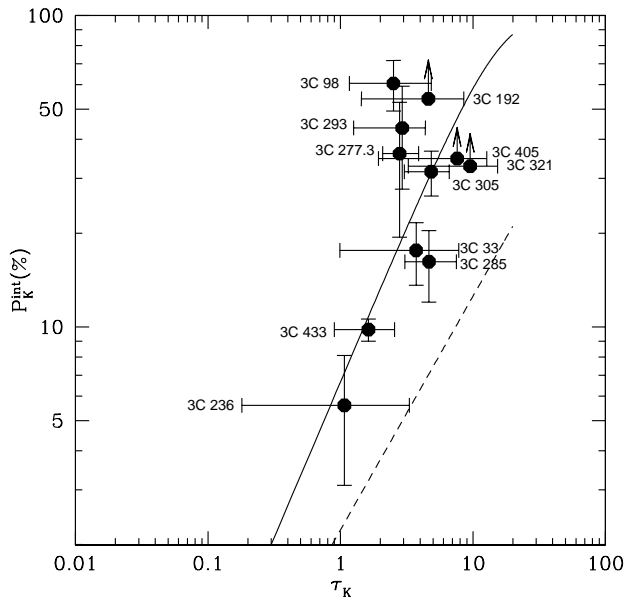}
\caption{The dichroic efficiency curve \citep{Jones:1989}. Solid and dashed lines represent the theoretical maximum and mean empirical dichroic efficiency respectively. The symbols represent the mean extinction $A_V$ estimated by independent methods for the 3CR radio galaxies \citep[see][]{Ramirez:2014}.}
\label{jones}
\end{figure}

Under the assumption that the intrinsic polarisation is due to the dichroic mechanism, the optical extinction vs. near-IR polarisation relation can be used \citep{Jones:1989} to deduce the $A_V$ required to obtain the measured $P^{\rm int}$. \citet{Jones:1989} took measurements of 2.2~$\mu$m polarisation of individual stars from the literature, and found a good correlation between the linear polarisation and the interstellar extinction. The stars that he analysed are extinguished by interstellar dust, and were chosen to be those without signs of intrinsic polarisation, or polarisation due to scattering.  Our near-IR polarisation estimates, and the mean optical extinction estimates for each object, are plotted on the dichroic efficiency curve of \citet{Jones:1989} in Fig. \ref{jones}. The filled circles in Fig. \ref{jones} represent the mean $A_V$ measurement estimated using the four different techniques in \citet[][excluding the derived optical extinction based on the silicate absorption line, which gives low extinction values]{Ramirez:2014}, and the error bars are the upper and lower values in the range of $A_V$ estimated using the four different techniques in \citet{Ramirez:2014}. According to \citet{Jones:1989}, Fig. \ref{jones} shows a clear trend for polarisation to be higher for larger optical depths. The trend he found can be fitted by a power-law (dashed line), with a slope of 0.75, following a relation given by:
\begin{equation}\label{Pmedian} 
P_K=2.23\tau_K^{3/4},
\end{equation}
where $\tau_K=0.09 A_V$ is the optical depth at K-band, derived using $A_{\lambda}=1.086\tau_{\lambda}$, and the extinction law $A_{\lambda}\propto\lambda^{-1.7}$ \citep{Mathis:1990}. The maximum polarisation percentage curve (solid line in Fig. \ref{jones}) is the upper envelope to the data following the relation $P_{max}=5.4 P_K$, where $P_K$ is the polarisation given in equation \ref{Pmedian}. In fact, this curve follows the theoretical relation given by \citep{Jones:1989}:
\begin{equation}\label{Pmax}
P_{max}=100\frac{(e^{\tau_p}-e^{-\tau_p})}{(e^{\tau_p}+e^{-\tau_p})},
\end{equation}
where $\tau_p=\tau_K (1-\eta)/(1+\eta)$, and $\eta$ is the dichroic efficiency of the dust grains: the ratio between the extinction perpendicular to the dust's long axis and the extinction parallel to the dust's long axis, typically in the range $0.87-0.93$. We set $\eta=0.875$ -- equivalent to the maximum interstellar polarisation found by \citet*{Serkowski:1975} and \citet{Jones:1989}.

From Fig. \ref{jones}, we find that, in general, our results are consistent with a dichroic origin for the polarisation only in the case of maximum dichroic efficiency. The exceptions are 3C~98, 3C~192, 3C~277.3 and 3C~293, for which even the case of maximum dichroic efficiency cannot reproduce the high level of intrinsic polarisation. It is worth mentioning that the correction for starlight for  3C~192 is the largest in the sample ($P^{\rm int}=60P$). The estimated dichroic extinction for 3C~433 \citep[see][]{Ramirez:2009} and 3C~405 \citep{Tadhunter:2000a} fits very well with the mean of the other extinction estimations. For the rest of the sources, the dichroic efficiency must be at the upper limit deduced for stars in the Milky Way, regardless of how the level of extinction is estimated. 

It is important to consider the uncertainty in the estimation of intrinsic polarisation. The sources of error are the polarisation measurement itself, and the estimation of the starlight dilution. In most of the cases the AGN contributes $<\!20$ per cent to the light in the nuclear aperture of 0.9 arcsec diameter (6 pixel radius). The starlight dilution is least in 3C~433 ($<\!30$ per cent). This source falls on the maximum dichroic efficiency curve in Fig. \ref{jones}, providing evidence for a dichroic origin of the polarisation \citep[see][]{Ramirez:2009}. Because the rest of the sources are consistent with 3C~433, this also provides evidence for a dichroic origin of the polarisation for the sources.

The relatively high dichroic efficiency can be explained in the following way: when we observe stars with large extinction in our Galaxy we may be observing them through several clouds, each with a different alignment of the dust grains, resulting in a low net polarisation. Most of the stars used to derive the dichroic efficiency curve by \citet{Jones:1989} could have the same problem. Hence, if the magnetic field in the torus is coherent, we could obtain higher efficiency than the average for stars in the Galaxy.

Furthermore, the $\textrm{PA}$s of the E-vectors provide information about both the dust properties and the magnetic fields in the torus \citep[e.g.][]{Packham:2007, Lopez-Rodriguez:2013}. The nuclear E-vector of the AGN is close to perpendicular to the radio axis in 4 out of 8 (50 per cent) sources in our complete {\em HST} sample, and in 6 out of 11 (54 per cent) sources in the extended sample (see Table \ref{tablepolarisation}). Given that dust grains tend to align their longest axis perpendicular to the magnetic field, this is evidence for the presence of a coherent toroidal magnetic field aligning elongated dust grains in the torus, leading to the measured orientation of the E-vector.

\subsection{Synchrotron radiation}

It remains the possibility that, at 2.05$~\mu$m, the flux may be contaminated by non-thermal radiation. It has been suggested that the intrinsic SEDs of the synchrotron core sources might follow a power-law that declines between the radio and the mid-IR wavelengths \citep{Dicken:2008}, or a parabolic shape\footnote[5]{The possibility of a parabolic SED shape, that is actually observed in some quasars \citep[e.g.][]{van_der_Wolk:2010}, cannot be tested for our sample objects because we do not have sufficient radio/sub-mm data to make an accurate parabolic fit.} \citep{Landau:1986,Leipski:2009,van_der_Wolk:2010}. By extrapolation to the near-IR it is then possible to determine whether the flux of the AGN at 2.05~$\mu$m is contaminated by synchrotron emission. 

In order to address this issue, it is necessary to examine the SEDs, and in particular the synchrotron radio core fluxes, to extrapolate to the near-IR. A power-law is assumed for this extrapolation. Measuring the slope between the $5$ GHz radio core measurements \citep{Giovannini:1988, Liuzzo:2009, Alexander:1984} and the millimetre (870$~\mu$m) wavelength measurements \citep*{Quillen:2003} for each radio galaxy, it is possible extrapolate to the near-IR wavelengths, comparing the extrapolated flux with the de-reddened 2.05$~\mu$m core fluxes, we can estimate the contamination from a non-thermal emission. 

\begin{figure*}
\leftline{\includegraphics[width=5.9cm,angle=0]{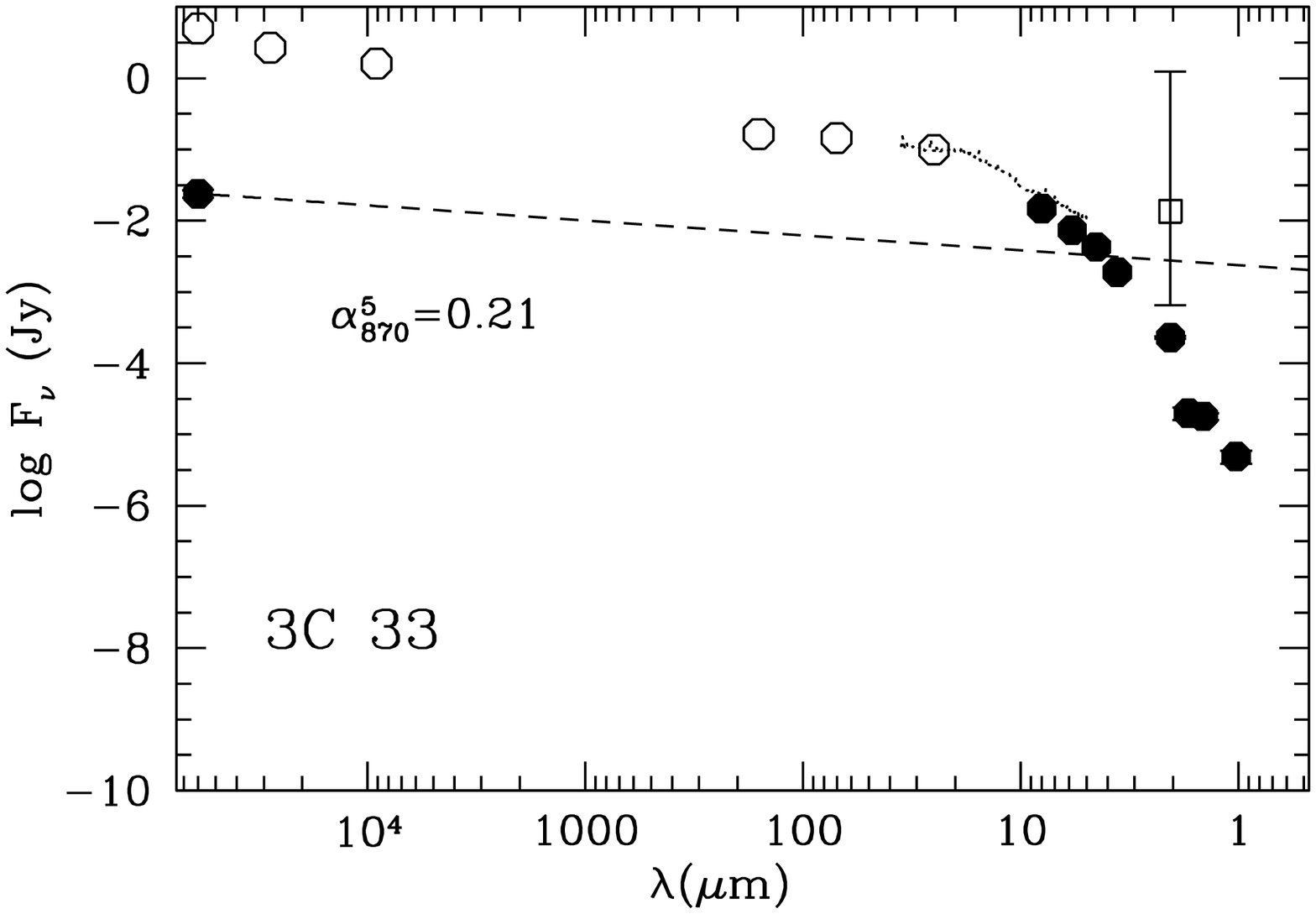}
\includegraphics[width=5.9cm,angle=0]{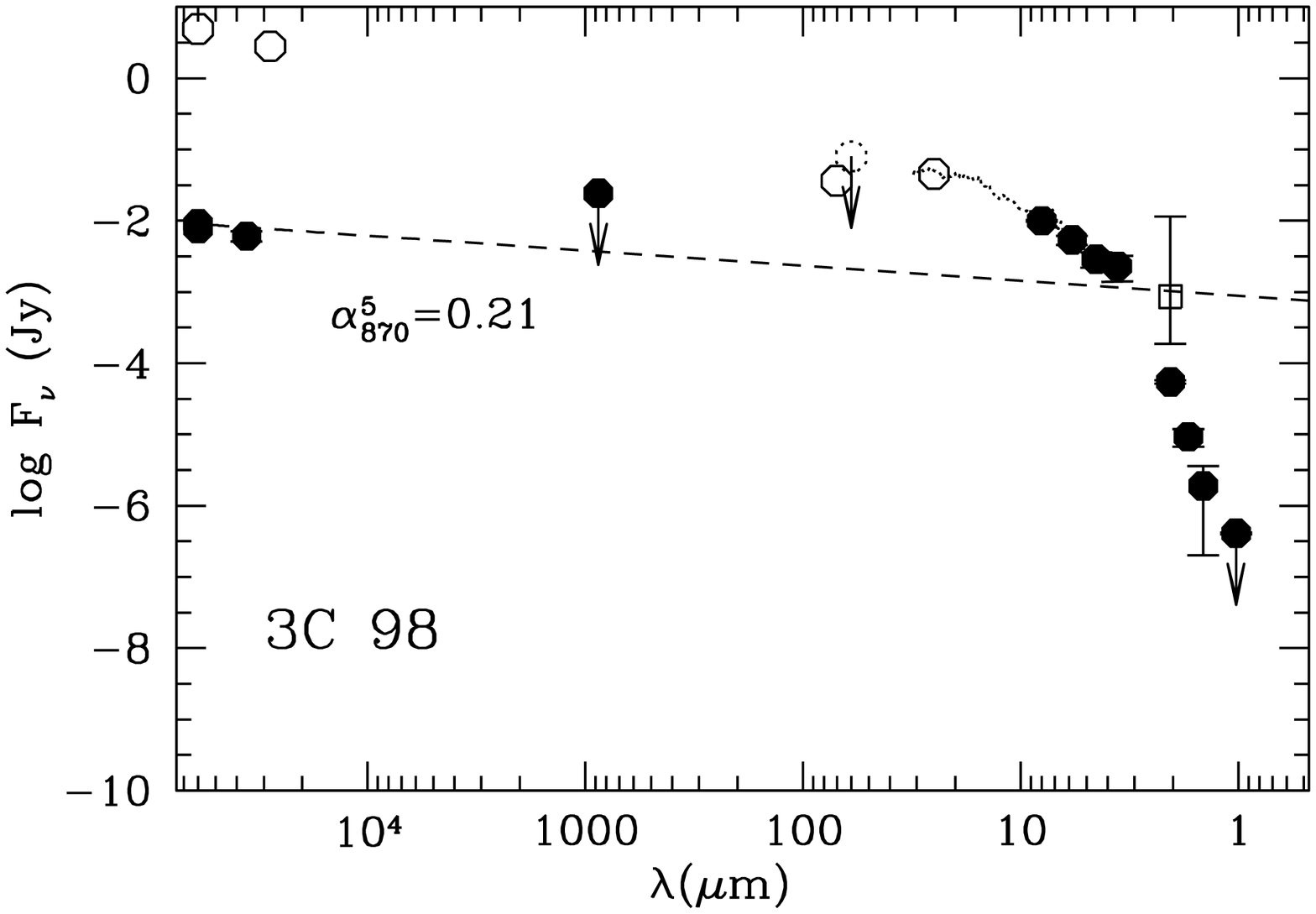}
\includegraphics[width=5.9cm,angle=0]{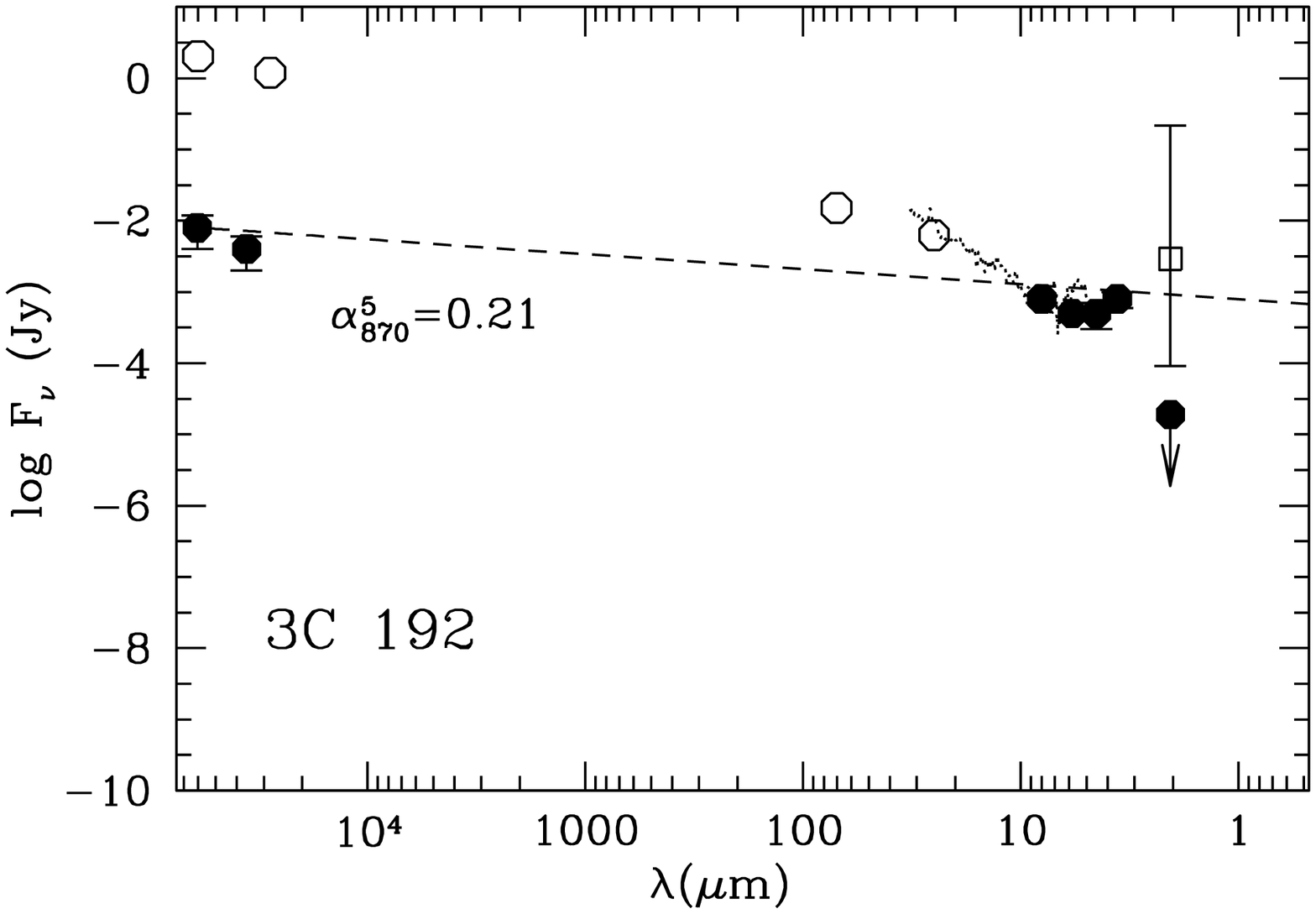}
}\vspace{-2cm}
\leftline{\includegraphics[width=5.9cm,angle=0]{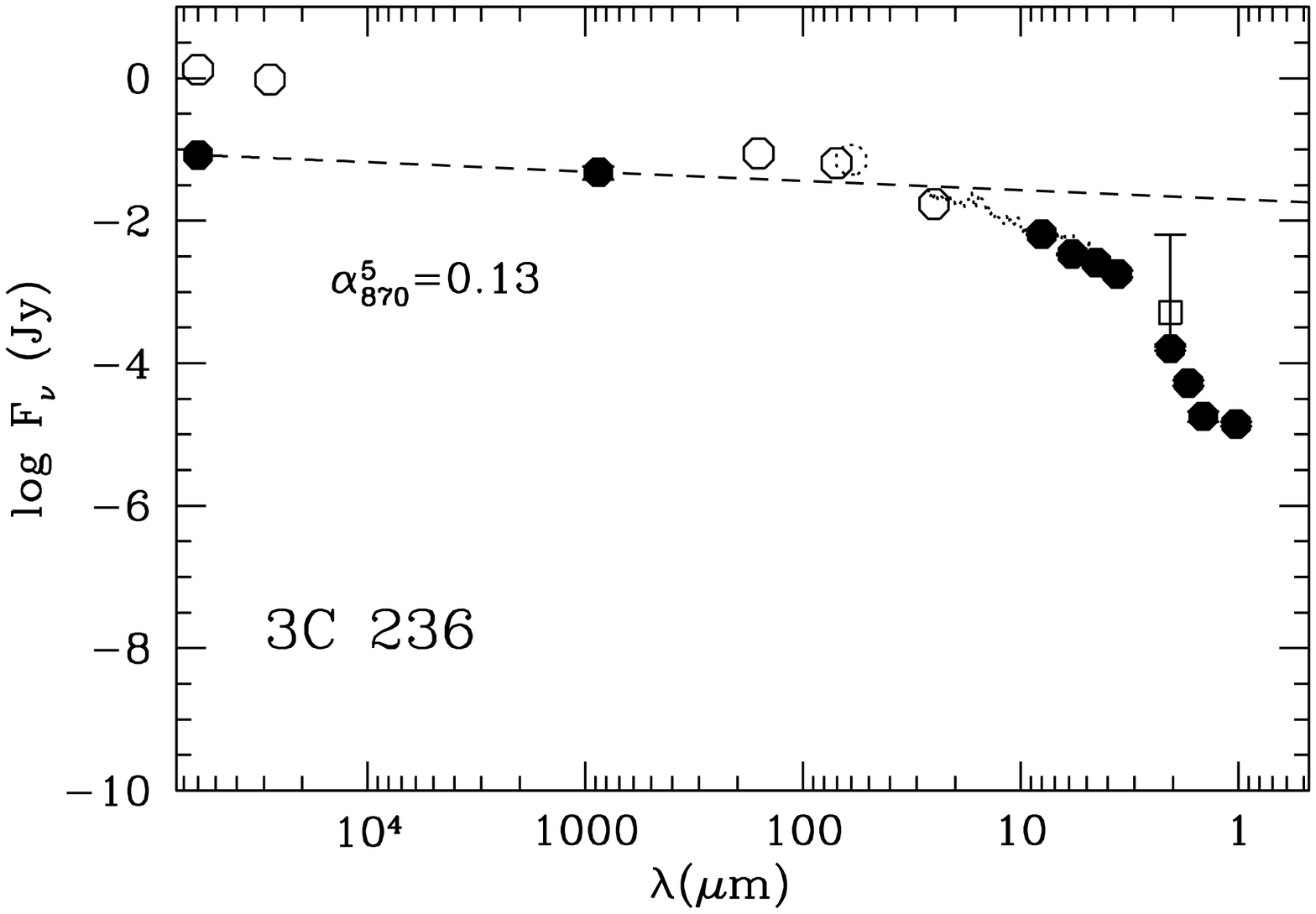}
\includegraphics[width=5.9cm,angle=0]{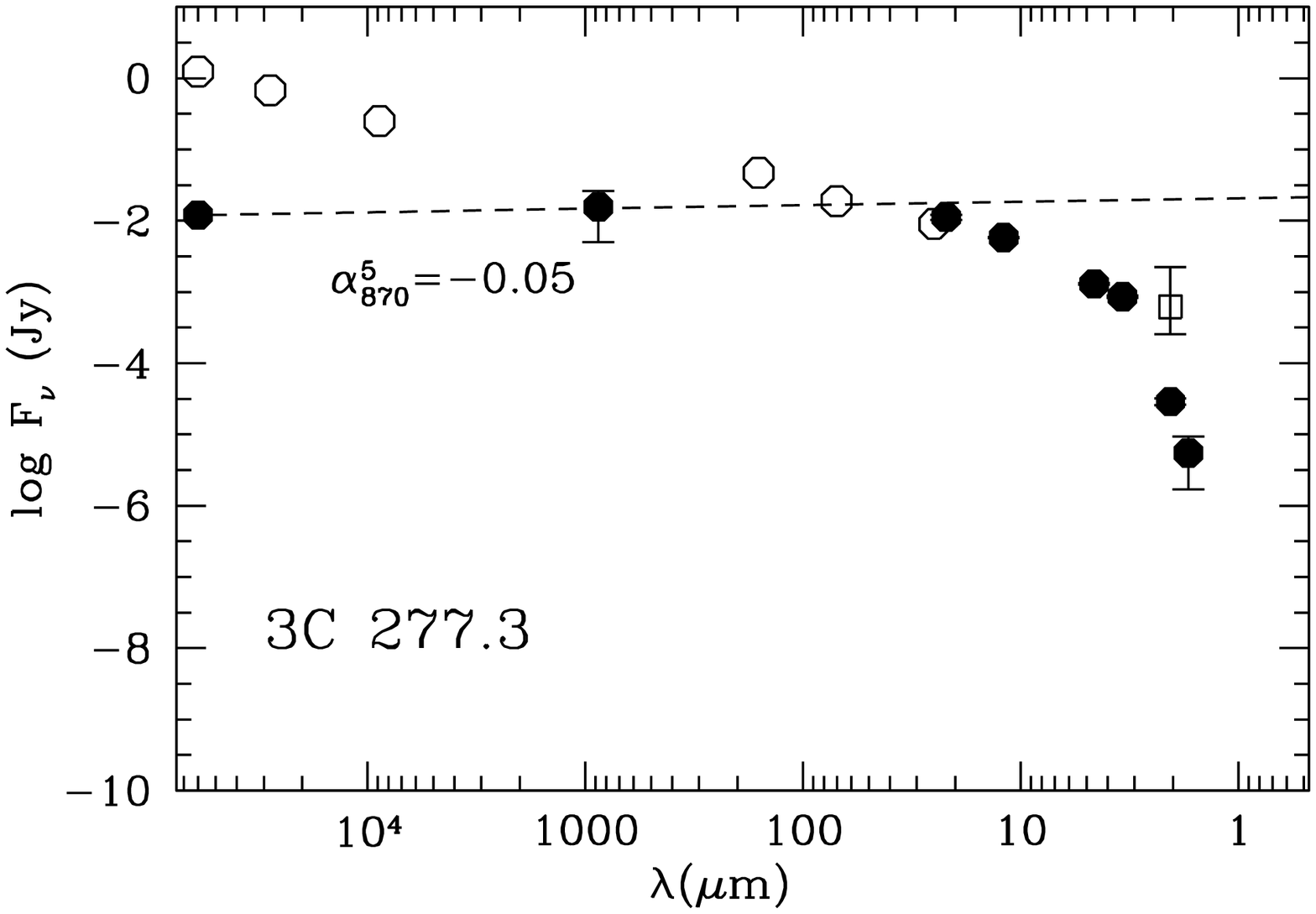}
\includegraphics[width=5.9cm,angle=0]{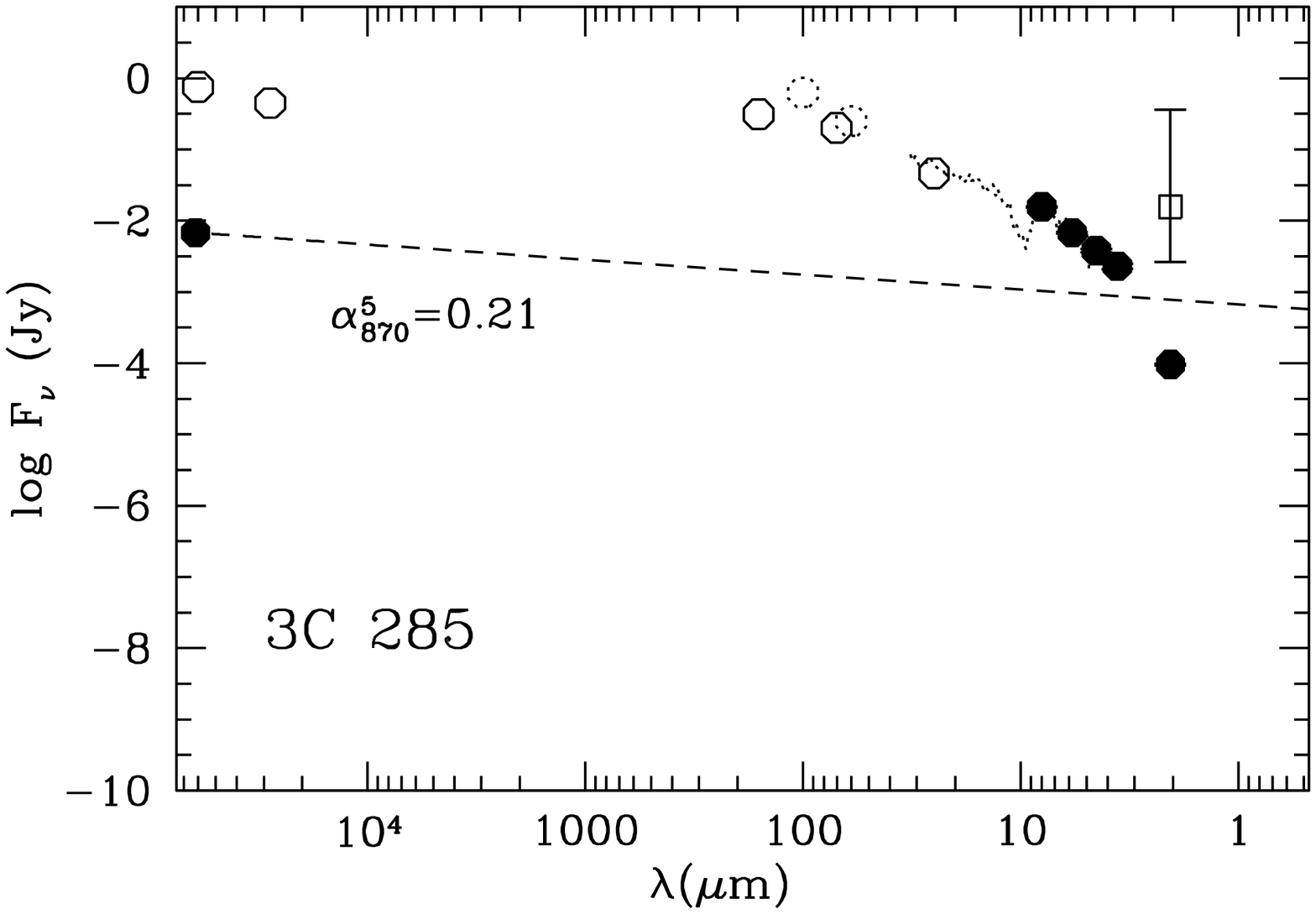}
}\vspace{-2cm}
\leftline{\includegraphics[width=5.9cm,angle=0]{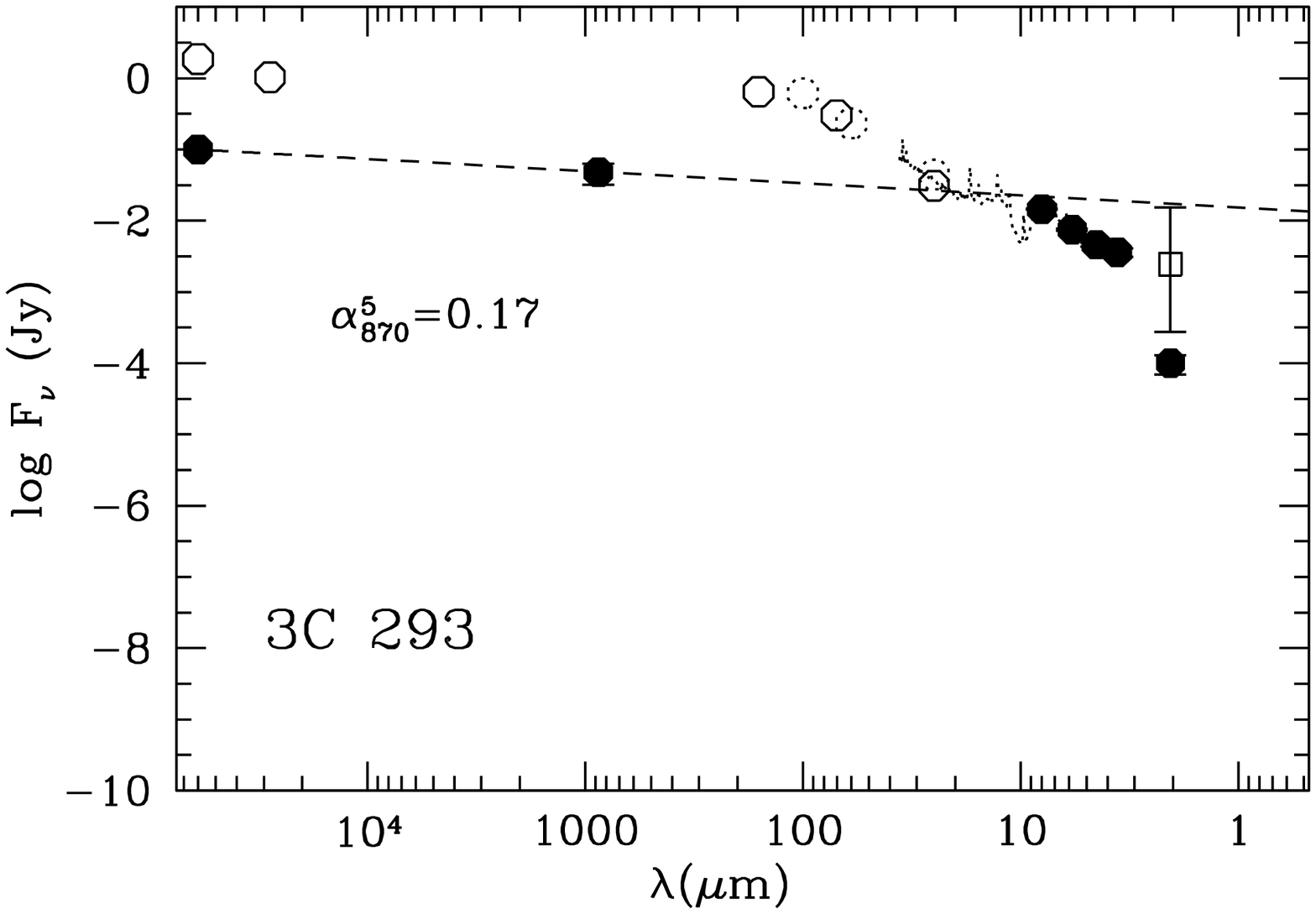}
\includegraphics[width=5.9cm,angle=0]{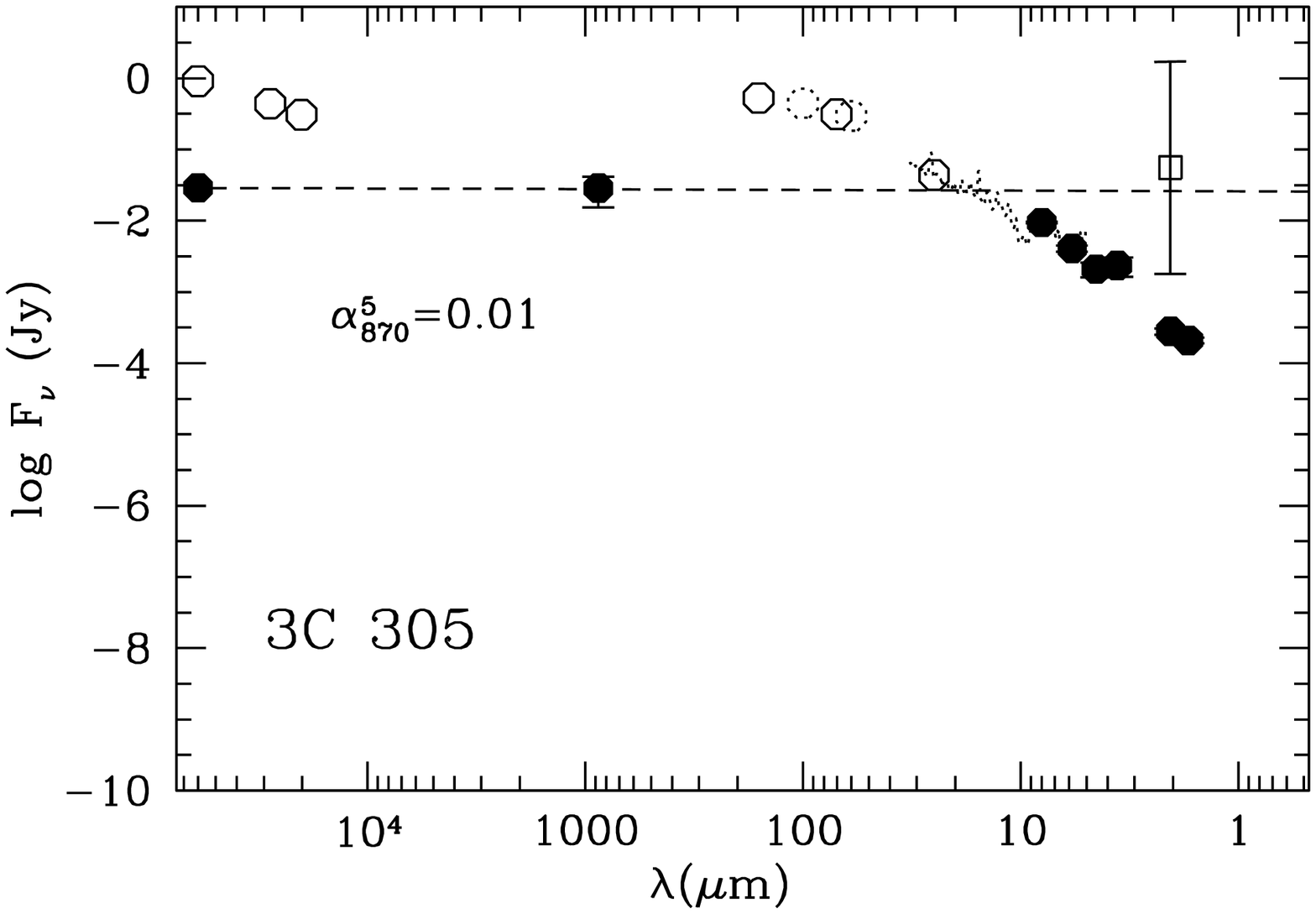}
\includegraphics[width=5.9cm,angle=0]{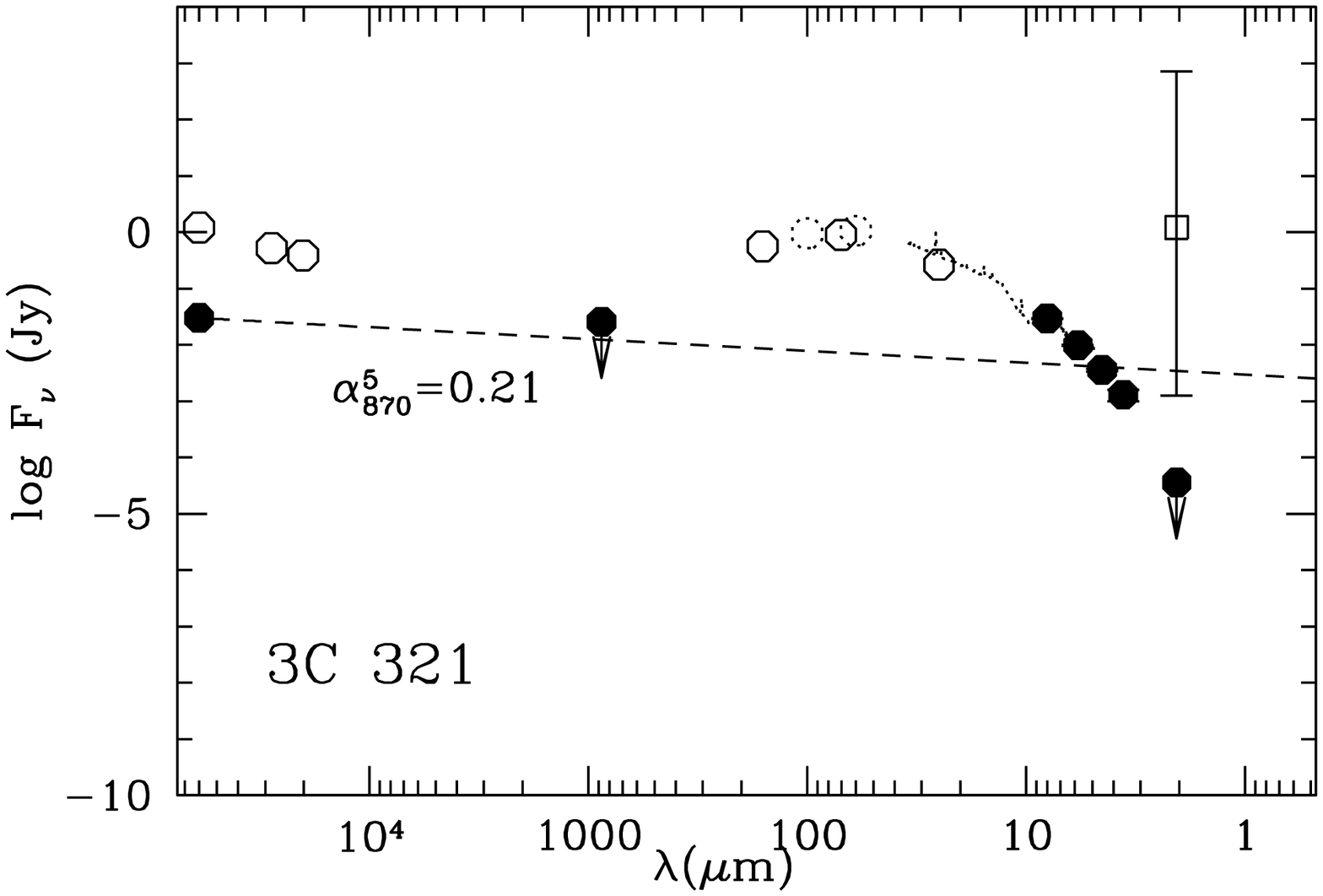}
}\vspace{-2cm}
\leftline{\includegraphics[width=5.9cm,angle=0]{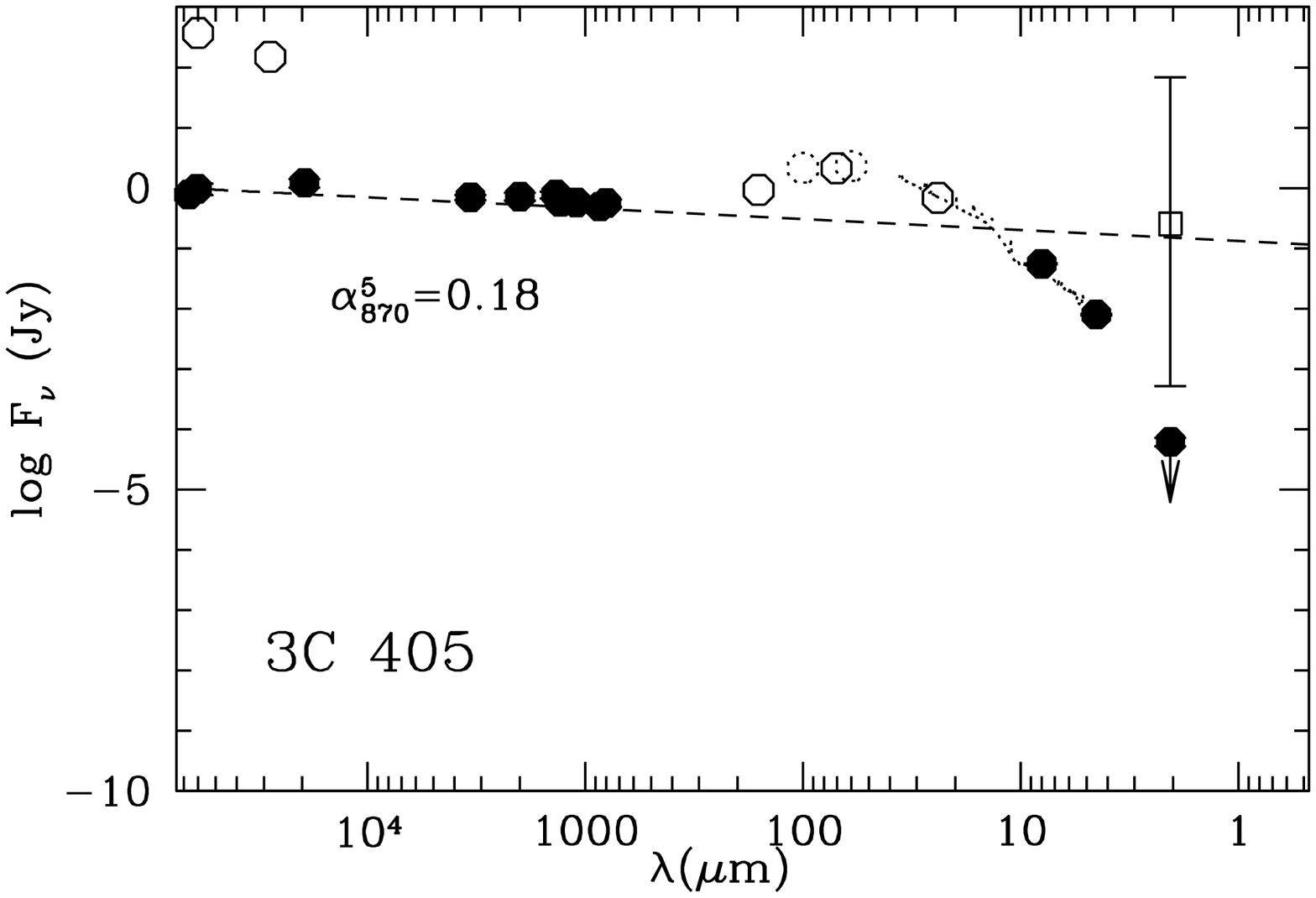}
\includegraphics[width=5.9cm,angle=0]{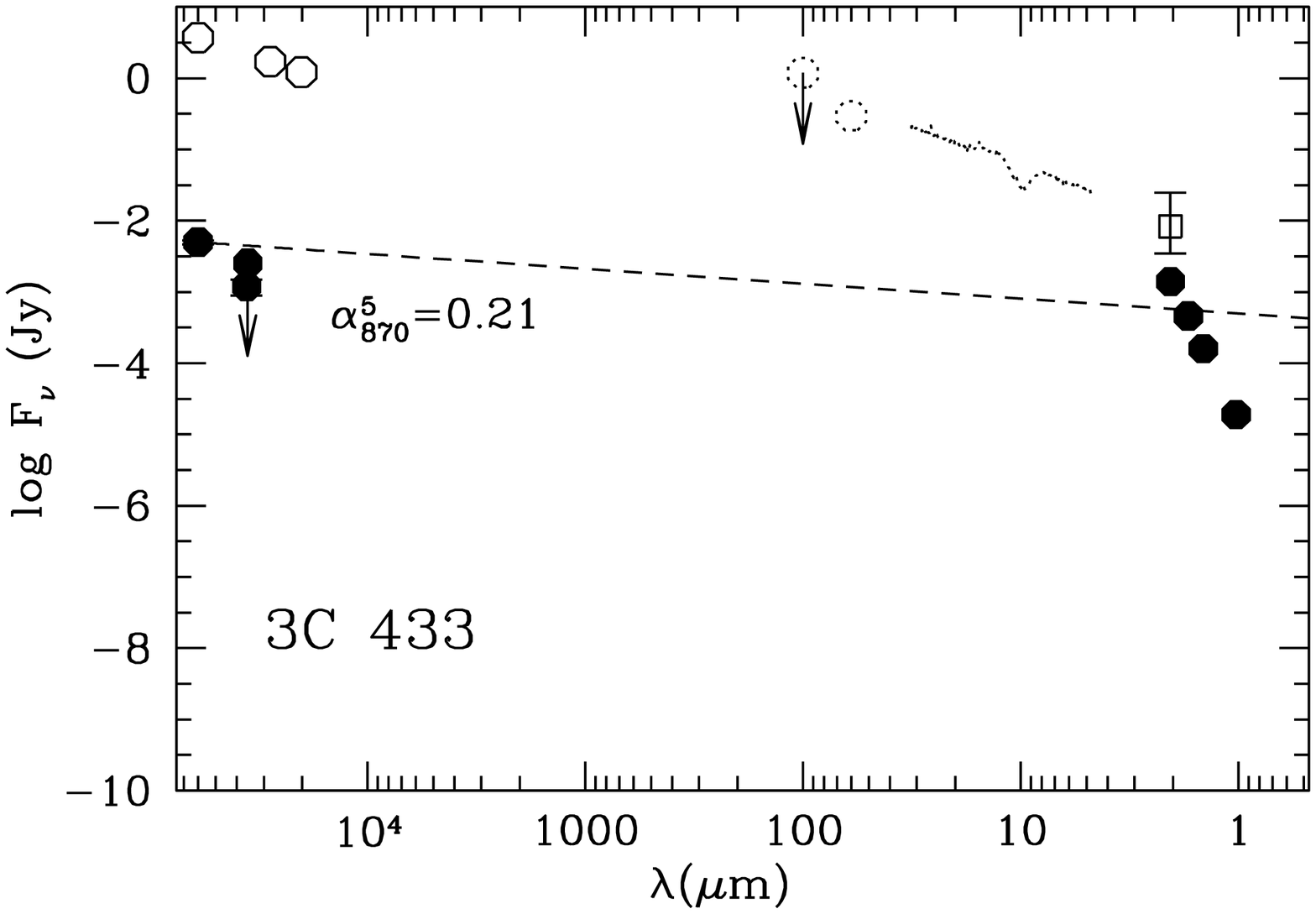}
}\vspace{-2cm}
\caption{SEDs of the radio galaxies in the extended sample from radio to the near-IR wavelengths. Filled circles represent the AGN core fluxes. Open squares represent the mean de-reddened 2.05~$\mu$m fluxes (using $A_V$ estimated by the different techniques in \citet{Ramirez:2014}, except the technique based on the silicate absorption line). The dotted line represents the IRS-{\em Spitzer} data \citep{Dicken:2012}. Open circles represent IRAC photometric measurements and the radio fluxes from the literature. Dashed open circles are the {\em Infrared Astronomical Satellite} ({\em IRAS}) mid-IR fluxes from the literature to compare with our measurements.  The dashed line is the power-law extrapolation from 5~GHz to the near-IR wavelengths. We have used the type-1 radio galaxies synchrotron power-law slope \citep[$\alpha^5_{870}=0.21$;][]{Quillen:2003} for cases in which this slope could not be directly measured. In cases that have actual measurements of the spectral index between the 5~GHz and millimetre wavelengths (870~$\mu$m), those were used for the extrapolation.}\label{tablefluxAGNsed}
\end{figure*}

The complete near- to far-IR SEDs including the radio and millimetric fluxes from the literature, are presented in Fig \ref{tablefluxAGNsed} \citetext{radio core fluxes at 5~GHz from \citealp{Giovannini:1988}, except 3C~277.3: \citealp{Liuzzo:2009}, and 3C~405: \citealp{Alexander:1984}}. Notice that there are sources in which a radio to millimetre slope had to be assumed because of the lack of millimetric core measurements (3C~33, 3C~192, 3C~285 and 3C~433), or because of the given upper limits to the millimetric flux (3C~98 and 3C~321). For these sources, the radio (5~GHz) to millimetre (870$~\mu$m) slopes were assumed to be equal to the slope deduced from  photometry of unobscured type-1 radio galaxies given in \citet{Quillen:2003}. \citet{Quillen:2003} used their millimetric continuum observations at 870~$\mu$m of a sample of 38 radio galaxies from the 3CRR sample \citep{Laing:1983}, and core fluxes at 5~GHz, to calculate the radio to millimetre spectral indices of 23 sources, of which 8 are type-1. Using these data, we estimate a median spectral index of $\alpha^5_{870}=0.21\pm0.1$ for type-1 radio sources.

In Fig. \ref{tablefluxAGNsed} we compare the power-law extrapolations with the mean de-reddened near-IR (2.05~$\mu$m) core measurement estimated by the four different techniques in \citet[][not including the method based on the silicate absorption line, which gives low extinction values]{Ramirez:2014}, to estimate the non-thermal contamination. Based on Fig. \ref{tablefluxAGNsed}, and taking into account the uncertainties in the de-reddened fluxes, it is possible that a non-thermal component makes a significant contribution at 2.05~$\mu$m in some sources. For instance, examining the extreme cases, the extrapolated flux for 3C~236, 3C~293 and 3C~277.3 falls far above the de-reddened near-IR core flux (see Fig. \ref{tablefluxAGNsed}), therefore, a significant contribution from non-thermal component is possible. On the other hand, 3C~285 and 3C~433 are expected to have no contribution from a polarised non-thermal component, given that the synchrotron extrapolation falls below the de-reddened near-IR flux. The other cases are not so clear.

A complication is that we do not know exactly what the intrinsic SED of the synchrotron emission is doing in the near-IR range of the spectra. Therefore the power-law extrapolation of radio and millimetric fluxes to near-IR wavelengths is a conservative approach for the following reasons.

\begin{itemize}
\item It is not clear where the non-thermal component turns over due to the ageing of the electron population. If the turnover occurs in the infrared, the non-thermal component will have a steeper slope from millimetre to near-IR wavelengths.
\item  Because of the possible parabolic shape (and the possible synchrotron ageing cut-offs), the simple power-law extrapolations are likely to provide an upper-limit on the synchrotron contribution at near-IR wavelengths. 
\item Synchrotron emission might not follow the same power-law at near-IR wavelengths. Actual measurements of the SEDs in some core-dominated Blazar sources (e.g. 3C~273), show a rapid decline by around two orders of magnitude from radio to near-IR wavelengths \citep{Dicken:2008}. 
\end{itemize}

Overall, with the exceptions of 3C~285 and 3C~433, we cannot rule out the idea that a non-thermal component produces the measured core polarisation. However, the measured position angles of the polarisation vectors at 2.05~$\mu$m, relative to the radio axis, are not what would be expected for synchrotron radiation. The polarisation vectors in the cores of radio galaxies \citep{Antonucci:1984,Lister:2001,Marscher:2002} tend to be parallel to the direction of the radio axis. In none of the sources in the extended sample is this the case (see Table \ref{tablepolarisation}): the offsets between the E-vectors measured at near-IR wavelengths and the radio jet axes are $>\!50\degree$ in 10 of the 11 cases with significant polarisation (the exception is 3C~293 with an offset of $29\pm11\degree$), and 6 are perpendicular to within $\pm20\degree$.

To further investigate the possibility that the non-thermal jet emission makes a significant contribution to the fluxes and polarisations of the near-IR core sources, we require measurements of the $\textrm{PA}$s of the polarisations in the radio cores of the individual sources in our sample. This will allow direct comparisons to be made with the near-IR core polarisations. Preferably the radio polarisation measurements should be made a high radio frequencies, in order to avoid the effects of Faraday rotation and depolarisation.

\subsection{Scattering}

Regarding scattering, the situation is more complicated because of the different possible scattering regions. The aperture size used to measure the core polarisation has a diameter of $0.9$ arcsec (6 pixels radius), equivalent to a physical size of $\sim 2$ kpc for $z\approx 0.1$ (the highest redshift of the sample). Within this aperture, scattering from several components could be present: from the far inner wall of the torus, from the outer parts of the torus, from an extended nebula \citep[like in 3C~405, where there is extended polarisation;][]{Tadhunter:2000a}, or from a reflecting cloud just above the torus. Because of the unresolved nature of the core sources, it is not possible to isolate particular components, making it difficult to rule out scattering as a mechanism.

For scattering, the polarisation angles are expected to be perpendicular to the radio axis in the torus wall and reflecting cloud cases, as observed in most objects (the $\textrm{PA}$ of the polarisation vector of 6 sources are perpendicular to within $\pm20\degree$). If the point sources are dominated by scattered AGN light, the actual AGN will be entirely hidden at near-IR wavelengths, implying that the extinction estimates from \citet{Ramirez:2014}, might substantially underestimate the true extinction to the AGN through the torus. 

In this context we note that a scattered near-IR component has been suggested by \citet{Baldi:2010} to explain the 1.6~$\mu$m flux excesses in NLRG-FRII galaxies. The near-IR (1.6 $\mu$m) core fluxes have been found to strongly correlate with the radio core fluxes \citep{Baldi:2010}: FRI objects follow a linear correlation, while broad-line FRII objects fall above the correlation. The near-IR excess in the broad-line FRII galaxies compared with the FRI and FRII classified as weak-line radio galaxies (WLRG) is attributed to the presence of direct emission from an accretion disc component. However, the near-IR core luminosities of the FRII NLRG fall somewhere between the near-IR core luminosities of FRI and broad-line FRIIs \citep{Baldi:2010}. This characteristic of the NLRG has been interpreted as being due to total obscuration of the direct light of the quasar at near-IR wavelengths, coupled with a significant scattered quasar component that enhances  the near-IR core luminosity.

\section{Discussion}\label{Discussion}

The nuclear and extended polarisations at 2.05~$\mu$m have been analysed for the sources in the complete and extended samples of radio galaxies. It has been found that the sources do not show any significant extended polarisation, apart from the well-known cases of 3C~293 \citep{Floyd:2006} and 3C~405 \citep{Tadhunter:2000a}. On the other hand, the nuclei are intrinsically highly polarised at near-IR wavelengths in most of the cases (except for 3C~452 and  4C~73.08).

We could be observing direct light from the AGN shining through a dusty torus, especially in those sources showing a clear presence of an unresolved point source at 2.05~$\mu$m \citep[in 10 out of 13 sources,][]{Ramirez:2014}, making it plausible that the polarisation in those sources is due to dichroic mechanism. If this is the case (polarisation by dichroic mechanism), then the dust grains must be aligned in a toroidal magnetic field in the torus. A toroidal magnetic field can have an important role in regulating the accretion of material feeding the black hole \citep*{Lovelace:1998}. Furthermore, a toroidal magnetic field could confine the torus' height against internal thermal pressure \citep{Lovelace:1998}. This would enable the existence of smaller and denser tori like those observed directly in interferometric observations \citep[e.g.][and references therein]{Elitzur:2008}. However, the polarisation could also be a result of a combination of the scattered flux and dichroic mechanisms, and it is impossible to distinguish the two components.

With additional polarimetry observations at other near- or mid-IR wavelengths, it might be possible to use the modified Serkowski curve \citep{Serkowski:1975} for the infrared range \citep[$P\propto \lambda^{-1.8}$;][]{Martin:1990}, to distinguish between the dichroic, scattering or synchrotron polarisation mechanisms. While dichroic extinction is strongly dependent on wavelength, following a $\lambda^{-1.8}$ relation, the intrinsic polarisation induced by scattering or synchrotron is only weakly dependent on wavelength. If the dichroic mechanism dominates, we can assume the empirical infrared curve $P\propto \lambda^{-1.8}$  \citep{Martin:1990}  to estimate the degree of polarisation, $P$, at 10~$\mu$m. In this way, we predict a mid-IR polarisation that is a factor of seventeen lower than the near-IR polarisation. For instance, we expect to measure polarisation at the $P\approx 0.5$ per cent level at 10~$\mu$m for the core of 3C~433. On the other hand, if scattered AGN light or synchrotron emission dominate across all near- to mid-IR wavelengths, we should expect negligible wavelength dependence of the polarisation in this wavelength range. In this case, the polarisation at 10~$\mu$m should be similar to that at 2.05~$\mu$m.

\section{Conclusion}\label{Conclusion}

In this paper the potential of near-IR polarimetric studies for investigating the structures of the circumnuclear tori in radio galaxies has been demonstrated. In particular, such studies have the potential to give us information about the orientation of the dust grains and hence the geometry of the magnetic field in the torus. The main conclusions of this paper are as follows. 

\begin{itemize}

\item After applying a correction for starlight, we find that the intrinsic polarisations of the compact near-IR sources of the radio galaxies in our sample ranges between $6$ and $60$ per cent. The nuclear polarisation is measured at the $>\!2\sigma$ in 11 out of 13 (85 per cent) for the {\em HST} extended sample, and at the $>\!3\sigma$ in 9 out of 13 (69 per cent) for the same sample. 

\item The extinction required by the maximum dichroic efficiency is consistent with the extinction estimated by the four different techniques discussed in \citet[][ not including the extinction estimates based on $\tau_{9.7}$, which gives low values to the optical extinction]{Ramirez:2014}. This consistency suggests that the high nuclear polarisation is due to dichroic extinction (except for 3C~293, which is more likely to have a non-thermal component), the product of elongated dust grains aligned in a magnetised torus in the nuclear region of the galaxy.

\item The $\textrm{PA}$s of the E-vectors in 6 out of 11 (54 per cent) sources in our {\em HST} sample, are perpendicular to the radio jet axis within $\pm20\degree$. The E-vectors are not aligned parallel to the radio axis in any case; the E-vector and the radio jet axis are offset by $>29\degree$ in all cases. This provides evidence for the presence of a coherent toroidal magnetic field aligning the dust grains.

\end{itemize}

In summary, we find that the 2.05~$\mu$m unresolved nuclei in some NLRG are intrinsically highly polarised (6 -- 60 per cent), with the E-vectors perpendicular to the radio axis in 54 per cent of the sources. If the polarisation is produced by dichroic extinction, then the mechanism must have its maximum efficiency, a product of dust grains aligned in a toroidal magnetic field in the torus.

\section*{Acknowledgments}

EAR thanks CONACyT and FAPESP. Based on observations made with the NASA/ESA Hubble Space Telescope, obtained from the data archive at the Space Telescope Science Institute (STScI). STScI is operated by the Association of Universities for Research in Astronomy, Inc. under NASA contract NAS 5-26555.

\bibliographystyle{mn2e}
\bibliography{bibliography}

\bsp

\label{lastpage}

\end{document}